\documentclass[showpacs,aps,floatfix,prd,preprintnumbers]{revtex4}
\usepackage{amssymb}
\usepackage{amsmath}

\usepackage{epstopdf}
\usepackage{hyperref}
\usepackage{capt-of}
\usepackage{graphicx}
\usepackage{dcolumn}
\usepackage{bm}
\usepackage{booktabs}
\usepackage{multirow}

\RequirePackage{float}

\begin{document}

\title{Accelerating Universe in Hybrid and Logarithmic Teleparallel Gravity}
\author{Sanjay Mandal$^{1}$\thanks{%
Email: sanjaymandal960@gmail.com}, Snehasish Bhattacharjee$^{2}$\thanks{%
Email: snehasish.bhattacharjee.666@gmail.com}, S. K. J. Pacif$^{3}$\thanks{%
Email: shibesh.math@gmail.com}, P.K. Sahoo$^{1}$\thanks{%
Email: pksahoo@hyderabad.bits-pilani.ac.in} }
\affiliation{$^{1}$ Department of Mathematics, Birla Institute of Technology and
Science-Pilani, Hyderabad Campus, Hyderabad-500078, India}
\affiliation{$^{2}$Department of Astronomy, Osmania University, Hyderabad-500007, India}
\affiliation{$^{3}$Department of Mathematics, School of Advanced Sciences, Vellore
Institute of Technology, Vellore 632014, Tamil Nadu, India.}

\begin{abstract}
Teleparallel gravity is a modified theory of gravity for which the Ricci
scalar $R$ of the underlying geometry in the action is replaced by an
arbitrary functional form of torsion scalar $T$. In doing so, cosmology in $%
f(T)$ gravity becomes greatly simplified owing to the fact that $T$ contains
only the first derivatives of the vierbeins. The article exploits this
appealing nature of $f(T)$ gravity and present cosmological scenarios from
hybrid and logarithmic teleparallel gravity models of the form $%
f=e^{mT}T^n $ and $f=D\log(bT)$ respectively, where $m$, $n$, $D$ and $b$
are free parameters constrained to suffice the late time acceleration. We
employ a well motivated parametrization of the deceleration parameter having
just one degree of freedom constrained with a $\chi^{2}$ test from 57 data
points of Hubble data set in the redshift range $0.07<z<2.36$, to obtain the
expressions of pressure, density and EoS parameter for both the teleparallel
gravity models and study their temporal evolution. We find the deceleration
parameter to experience a signature flipping for the $\chi^{2}$ value of the
free parameter at $z_{tr}\simeq0.6$ which is consistent with latest Planck
measurements. Next, we present few geometric diagnostics of this
parametrization to understand the nature of dark energy and its deviation
from the $\Lambda$CDM cosmology. Finally, we study the energy conditions to
check the consistency of the parameter spaces for both the teleparallel
gravity models. We find the SEC to violate for both the models which is an
essential recipe to obtain an accelerating universe.
\end{abstract}

\maketitle

%\input epsf.tex
%%%%%%%%%%%%
%%%%%%%%%%%

\section{Introduction}\label{I}

Several Observations reveal that the universe is accelerating for the very
second time in its 13.7 billion year long lifetime \cite{observations}. It
has now been agreed that a cosmological entity with almost three-quarters of
the energy budget of the universe coupled with a EoS parameter $%
\omega\simeq-1$ is required to suffice the observations. In this spirit,
several interesting proposals have been reported to expound this conundrum 
\cite{alternate}. \newline
One of the most interesting proposal refuting the existence of dark energy
are the `modified theories of gravity'. In modified gravity theories, dark
energy is purely geometric in nature and is connected to novel dynamical
terms following modification of the Einstein-Hilbert action \cite{cosmoft}.
Many such theories such as $f(R)$ gravity, $f(G)$ gravity, $f(R,T)$ gravity,
etc have widespread use in modern cosmology (For a recent review on modified
gravity see \cite{review}. Also see \cite{snehasish} for some interesting cosmological applications of modified gravity). \newline
Teleparallel gravity is a well established and well motivated modified
theory of gravity inspired from $f(R)$ gravity \cite{cosmo7to9} (See \cite%
{reviewft} for a review on teleparallel gravity). In teleparallel gravity,
the Ricci scalar $R$ of the underlying geometry in the action is replaced by
an arbitrary functional form of torsion scalar $T$. Thus, in teleparallel
gravity, instead of using the torsionless Levi-Civita connection (which is
usually assumed in GR), the curvatureless Weitzenb\"{o}ck connection is
employed in which the corresponding dynamical fields are the four linearly
independent verbeins, and $T$ is related to the antisymmetric connection
following from the non-holonomic basis \cite{cosmoft,cosmo10to12}.\newline
Linear $f(T)$ gravity models are the teleparallel equivalent of GR (TEGR) 
\cite{cosmo14}. Nonetheless, $f(T)$ gravity differ significantly from $f(R)$
gravity in the fact that the field equations in $f(T)$ gravity are always at
second-order compared to the usual fourth-order in $f(R)$ gravity. This owes
to the fact that the torsion scalar contains only the first derivatives of
the vierbeins and thus makes cosmology in $f(T)$ gravity much simpler.
However, Despite being a second-order theory, very few exact solutions of
the field equations have been reported in literature. Power law solutions in
FLRW spacetime have been reported in \cite{cosmo19to20}, while for
anisotropic spacetimes in \cite{cosmo21}. Solutions for Bianchi I spacetime
and static spherically spacetimes can be found in \cite{cosmoft} and \cite%
{cosmo22to23} respectively. \newline
Since cosmology in $f(T)$ gravity is much simpler compared to other modified
gravity theories, it has been employed to model inflation \cite{inflation},
late time acceleration \cite{late} and big bounce \cite{bounce}. The
instability epochs of self-gravitating objects coupled with anisotropic
radiative matter content and the instability of cylindrical compact object
in $f(T)$ gravity have been discussed in Ref. \cite{bhatti/2017,bhatti/2017a}%
. \newline
The manuscript is organized as follows: In Section \ref{II} we present an
overview of $f(T)$ gravity. In Section \ref{III} we describe the kinematic
variables obtained from a parametrization of deceleration parameter used to
obtain the exact solutions of the field equations. In Section \ref{IV} we
present the hybrid and logarithmic teleparallel gravity models and obtain
the expressions of pressure, density and EoS parameter. In Section \ref{V}
we present some geometric diagnostics of the parametrization of deceleration
parameter. In Section \ref{VI} we study the energy conditions for both the
teleparallel gravity models. In Section \ref{VII} we obtain some
observational bounds on the free parameters of the parametrization by
performing a chi-square test using Hubble datasets with $57$ datapoints,
Supernovae datasets consisting of $580$ data points from Union$2.1$
compilation datasets and Baryonic Aucostic Oscillation (BAO) datasets. Finally, in Section \ref{VIII} we present our
results and conclusions.

\section{Overview of $f(T)$ Gravity}\label{II}

The action in teleparallel gravity is represented as 
\begin{equation}  \label{1a}
S=\frac{1}{16\pi G}\int[T+f(T)]e d^4x,
\end{equation}
where $e=det(e^i_\mu)=\sqrt{-g}$ and $G$ is Newtonian gravitational
constant. The gravitational field in this framework arises due to torsion
defined as 
\begin{equation}  \label{1b}
T^{\gamma}_{\mu \nu}\equiv e^{\gamma}_i(\partial_{\mu}
e^i_{\nu}-\partial_{\nu} e^i_{\mu}).
\end{equation}
The contracted form of torsion tensor reads 
\begin{equation}  \label{1c}
T\equiv \frac{1}{4}T^{\gamma \mu \nu}T_{\gamma \mu \nu}+\frac{1}{2}T^{\gamma
\mu \nu}T_{\nu \mu \gamma}-T^{\gamma}_{\gamma \mu}T^{\nu \mu}_{\nu}.
\end{equation}
varying the action $S+L_m$, where $L_m$ represent the matter Lagrangian
yields the field equations as 
\begin{multline}  \label{1d}
e^{-1}\partial_{\mu}(ee^{\gamma}_i S^{\mu
\nu}_{\gamma})(1+f_T)-(1+f_T)e^{\lambda}_i T^{\gamma}_{\mu \lambda}S^{\nu
\mu}_{\gamma} +e^{\gamma}_i S^{\mu \nu}_{\gamma}\partial_{\mu}(T)f_{TT}+%
\frac{1}{4}e^{\nu}_i[T+f(T)]=\frac{k^2}{2} e^{\gamma}_iT^{(M)\nu}_{\gamma},
\end{multline}
where $f_T=df(T)/dT$, $f_{TT}=d^2f(T)/dT^2$, the ``superpotential" tensor $%
S^{\mu \nu}_{\gamma}$ written in terms of cotorsion $K^{\mu \nu}_{\gamma}=-%
\frac{1}{2}(T^{\mu \nu}_{\gamma}-T^{\nu \mu}_{\alpha}-T^{\mu \nu}_{\alpha})$
as $S^{\mu \nu}_{\gamma}=\frac{1}{2}(K^{\mu
\nu}_{\gamma}+\delta^{\mu}_{\gamma}T^{\alpha
\nu}_{\alpha}-\delta^{\nu}_{\gamma}T^{\alpha \mu}_{\alpha})$ and $%
T^{(M)\nu}_{\gamma}$ represents the energy-momentum tensor to the matter
Lagrangian $L_m$. For a flat FLRW universe with the metric denoted as 
\begin{equation}  \label{1e}
ds^2=dt^2-a^2(t)dx^{\mu} dx^{\nu},
\end{equation}
where $a(t)$ the scale factor, gives 
\begin{equation}  \label{1f}
e^{i}_{\mu}=diag(1,a,a,a).
\end{equation}
Employing \eqref{1e} into the field equation \eqref{1d}, the modified
Friedman equations reads 
\begin{equation}  \label{4a}
H^{2}=\frac{8 \pi G}{3}\rho -\frac{f}{6} + \frac{T f_{T}}{3},
\end{equation}
\begin{equation}  \label{4b}
\dot{H}=-\left[\frac{4 \pi G (\rho+p)}{1+f_{T}+2 T f_{T T}} \right] ,
\end{equation}
where $H\equiv \dot{a}/a$ denote the Hubble parameter and dots represent the
derivative with respect to time and $\rho$ and $p$ be the energy density and
pressure of the matter content and $T=-6H^{2}$. From equations \eqref{4a}
and \eqref{4b}, we obtain the expressions of density $\rho$, pressure $p$
and EoS parameter $\omega$ respectively as

\begin{equation}  \label{rho}
\rho=3H^2 +\frac{f}{2} +6H^2f_T
\end{equation}
\begin{equation}  \label{pre}
p=-2\dot{H}(1 + f_T -12H^2f_{TT}) - (3H^2 +\frac{f}{2}+6H^2f_T)
\end{equation}
\begin{equation}  \label{omega}
\omega=\frac{p}{\rho}=-1-\frac{2\dot{H}(1 + f_T -12H^2f_{TT})}{(3H^2 +\frac{f%
}{2}+6H^2f_T)}
\end{equation}
where we set $8\pi G=1$. Furthermore, the continuity equation reads 
\begin{equation}
\dot{\rho}+3H(1+\omega)\rho=0,
\end{equation}

\section{Kinematic Variables}\label{III}

The system of field equations described above has only two independent
equations with four unknowns. To solve the system completely and in order to
study the temporal evolution of energy density, pressure and EoS parameter,
we need two more constraint equations (extra conditions). In literature,
there are several arguments to choose these equations (see \cite{pacif} for
details). The method is well known as the model independent way approach to
study cosmological models that generally considers a parametrizations of any
kinematic variables such as Hubble parameter, deceleration parameter, jerk
parameter and EoS parameter and provide the necessary supplementary equation 
\cite{para}. Bearing that in mind, we shall work with a parametrization of
deceleration parameter proposed in \cite{banerjee} as 
\begin{equation}
q=-1+\alpha \left[ -1+\frac{1}{1+\left( \frac{1}{1+z}\right) ^{\alpha }}%
\right]   \label{5e}
\end{equation}%
where $\alpha $ shall be constrained from a chi-square test using any
observational datasets (ref. section VII). The motivation use this
parametrization is driven by the fact that equation \eqref{5e} allows a
signature flipping for $-1>\alpha >-2$. Additionally note that

\begin{itemize}
\item $\alpha=-2$ corresponds to a decelerated universe at $z=0$.

\item $\alpha<-2$ corresponds to an accelerated universe in the future
(i.e., $z<0$).

\item $\alpha\geq-1$ corresponds to an externally accelerating universe.
\end{itemize}

The expression of Hubble parameter for the parametrization \eqref{5e} reads 
\begin{equation}  \label{5d}
H=\beta \left[1+\left(\frac{1}{1+z}\right)^ {\alpha}\right]
\end{equation}
where $\beta$ is the integration constant. To obtain \eqref{5d}, we used the
relation 
\begin{equation}
\frac{H (z)}{H_{0}}=\exp\left[\int^{z}_{0}\frac{1+ q(z^{\prime })}{%
1+z^{\prime }}dz^{\prime }\right]
\end{equation}
%From the above equation of $q$, we find the range of the deceleration parameter $q \in [(-1-\alpha),-1]$. At $z=0$, the value of deceleration parameter is given by $q_0=-1-\frac{\alpha}{2}$ which is called as the present value of $q(z)$. For, $\alpha<-1$, the signature of the deceleration parameter changes from positive to negative. Therefore, we are interested to study the evolution of the Universe as suggested by the observations \cite{Perlmutter/1999,Riess/1998,stern/2010}. For that we are restricted the parameter $\alpha (-2<\alpha<-1)$ in our model. We plotted the deceleration parameter $q(z)$ for  different values of $\alpha$ in the range $(-2<\alpha<-1)$ to discuss it's behaviour as shown in Fig. \ref{f1} . For other values of $\alpha$, we discussed as follows

\begin{figure}[H]
\centering
\includegraphics[width=8.5 cm]{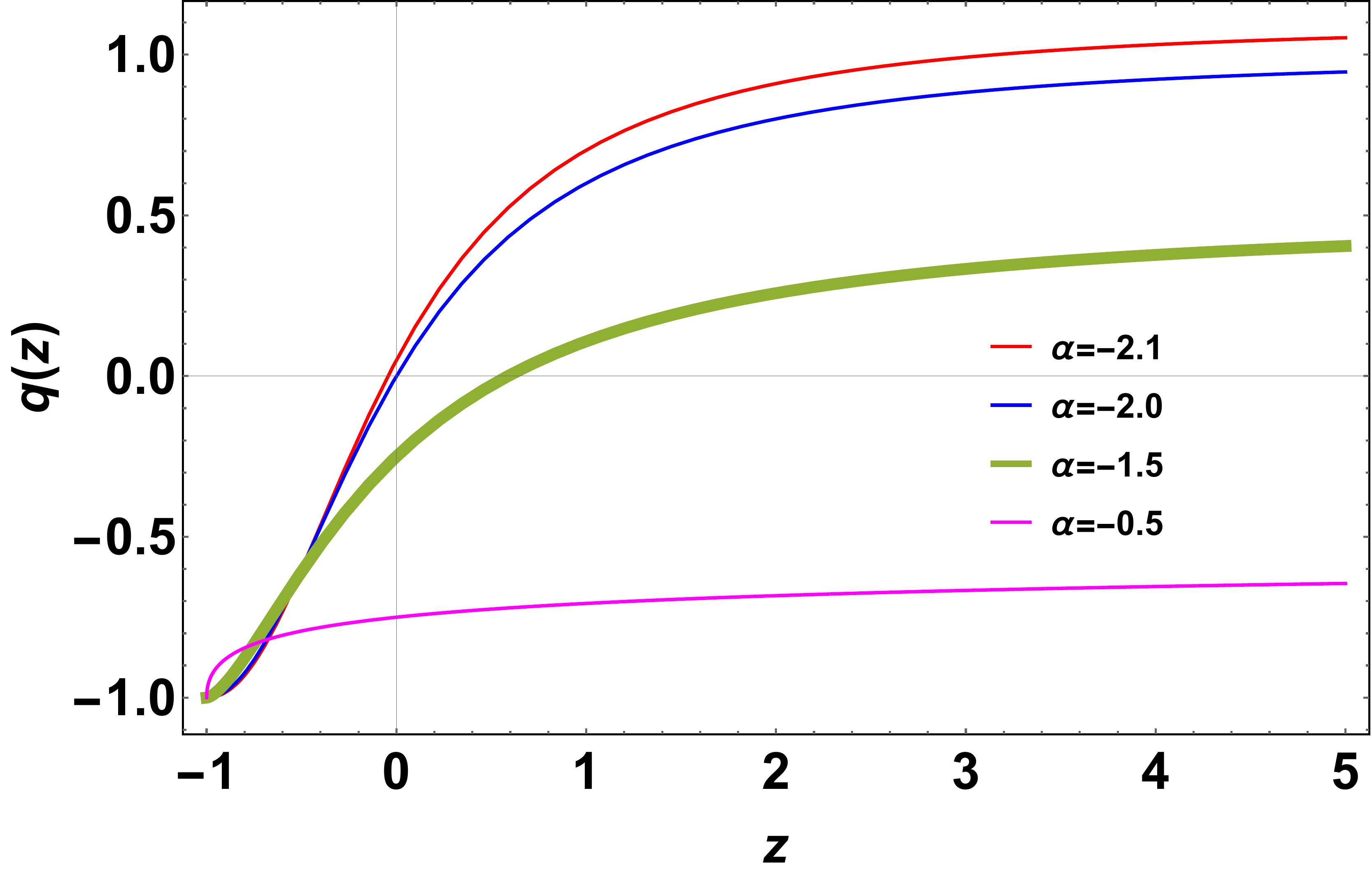}
\caption{Plot of deceleration parameter ($q$) as a function of redshift ($z$%
) for different values of model parameter $\protect\alpha $ showing diverge
evolutionary dynamics.}
\label{f1}
\end{figure}
%The cosmological  ``phase transition" act as an important role to study the evolution of the universe i.e the moment when there is no acceleration and deceleration in the universe. In Fig.\ref{f1}, we have seen that the phase transition happens in $(-2,-1)$ range of model parameter $\alpha$. Clearly, it is observed that for $\alpha=-1.56, \alpha=-1.449729, \alpha=-1.42$ the phase transition happens at $0.74, 0.59, 0.52$, respectively, for which the behaviour of deceleration parameter $q$ obeys the  observational data\cite{Santos/2006,Gong/2006} as shown in Fig.\ref{f1}.\\
%The geometrical parameter always act a significant role to study and get more information about the universe. Therefore, here we have discussed the geometrical parameters by taking higher order derivatives of scale factor. The Hubble parameter ($H$), deceleration parameter ($q$), jerk ($j$), snap ($s$), and lerk ($l$) are first five derivatives of scale factor, respectively. All the parameters are dimensionless except the Hubble parameter which is inverse of time. The geometrical parameters can be written as follows

\begin{figure}[H]
\centering
\includegraphics[width=8.5 cm]{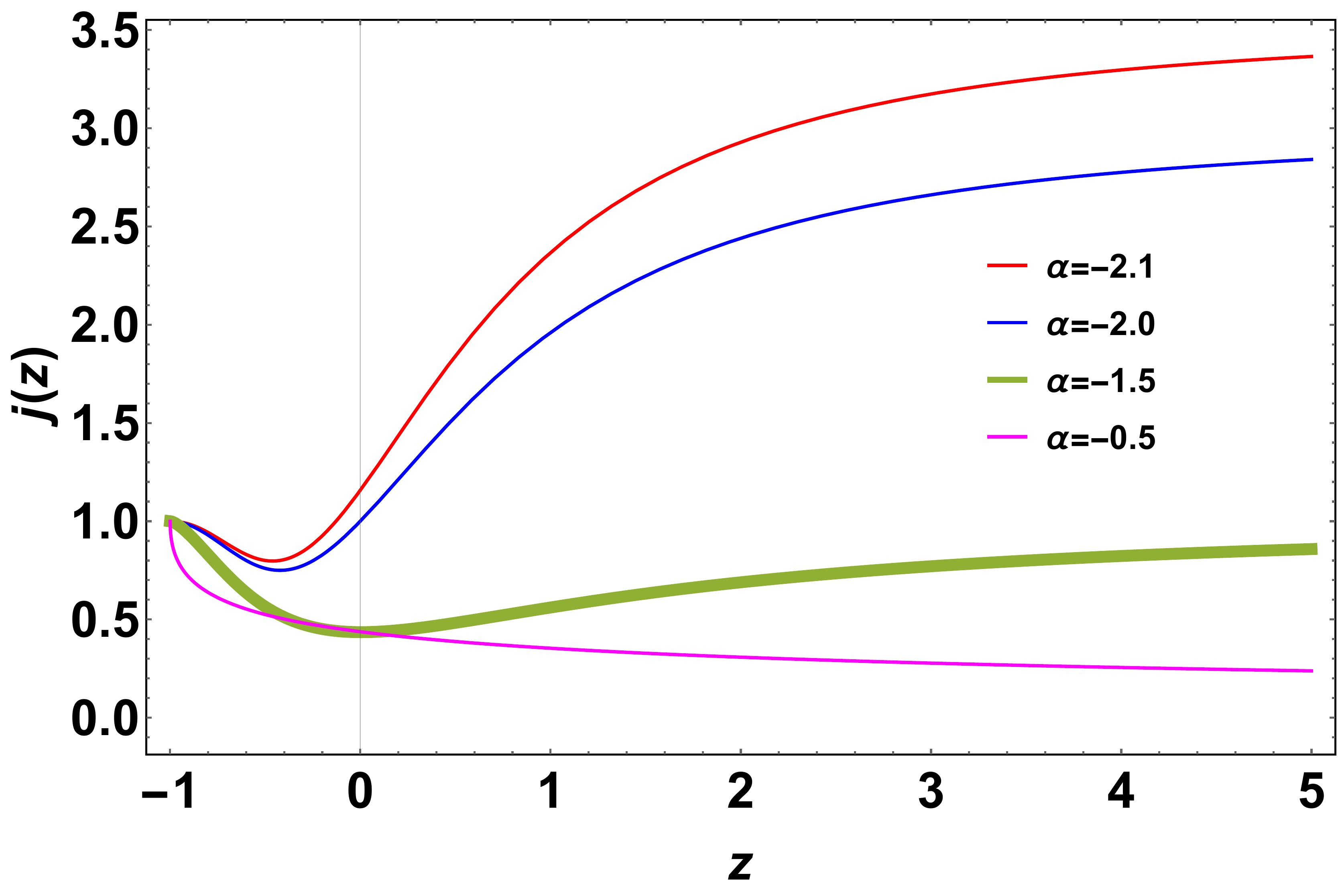}
\caption{Jerk parameter as a function of redshift.}
\label{f1a}
\end{figure}

\begin{figure}[H]
\centering
\includegraphics[width=8.5 cm]{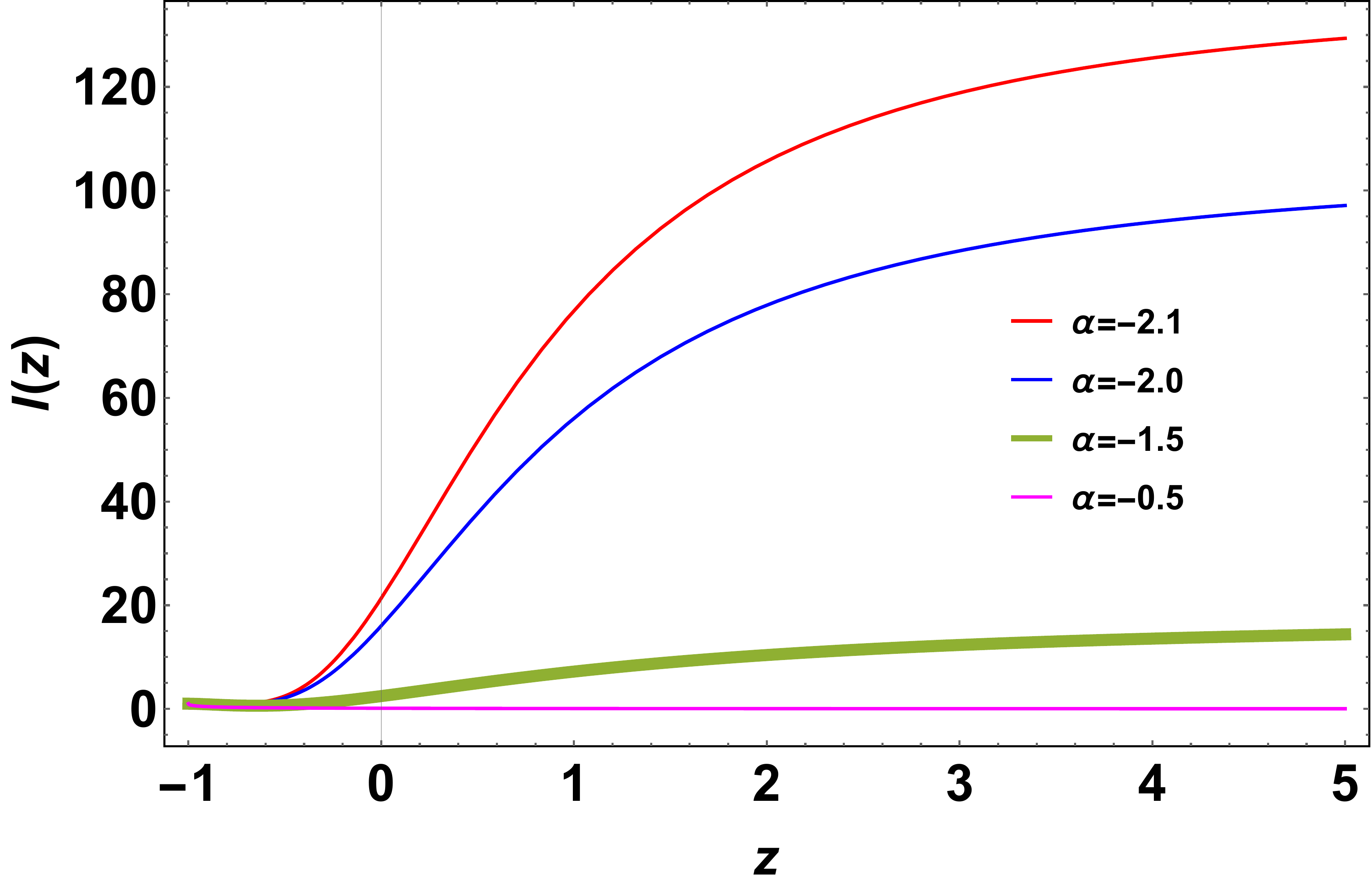}
\caption{Lerk parameter as a function of redshift.}
\label{f1c}
\end{figure}

Higher order derivatives of deceleration parameter such as jerk ($j$), snap (%
$s$) and lerk ($l$) parameters provide important information about the
evolution of the universe. They are represented as \cite{olive21} 
\begin{align*}
j(z)=(1+z)\frac{dq}{dz}+q(1+2q),
\end{align*}
\begin{align*}
s(z)=-(1+z)\frac{dj}{dz}-j(2+3q),
\end{align*}
\begin{align*}
l(z)=-(1+z)\frac{ds}{dz}-s(3+4q)
\end{align*}
The jerk parameter represents the evolution of deceleration parameter. Since 
$q$ can be constrained from observations, jerk parameter is used to predict
the future. Additionally, the jerk parameter along with higher derivatives
such as snap and lerk parameters provide useful insights into the emergence
of sudden future singularities \cite{olive21}.\newline
From Fig. \ref{f1a} and Fig. \ref{f1c}, the jerk and lerk parameters are
observed to have decreasing behaviors. Also, as the value of $\alpha$
decreases, the parameters assumes higher values at redshift $z=0$. Both of
these parameters are positive which represents an accelerated expansion. The
snap parameter is negative for all $\alpha$ which also denote an accelerated
expansion. Interestingly, the jerk parameter does not attain unity at $z=0$
which clearly does not coincide with $\Lambda$CDM model. Interestingly, this
implies that the late time acceleration can be caused due to modifications
of gravity. It is therefore encouraging to study the dynamics of EoS
parameter which may arise purely due to geometric effects in the framework
of modified gravity theories such as teleparallel gravity.

%The jerk parameter has a great feature i.e. it is equivalent to $\Lambda$CDM model for $j=1$, which is noted in \cite{Sahani/2003,Alam/2003}. Any deviation from the value $j=1$ defines the dark energy model which is different from $\Lambda$CDM model. The parametrization of jerk parameter is an alternative approach to describe the cosmological model which is close to agreement of $\Lambda$CDM model. The parametrization of jerk parameter has been presented in \cite{Zhai/2013,Mukherjee/2016,Mamon/2018,Capozziello/2020}.In Fig.\ref{f1a}, the profile of $j$ has been presented throughout the evolution of time and it is clearly observed the sign of $j$ doesn't change. Since $q$ is observed, and $j$ represents the time variation of the deceleration parameter. So, $j$ is used to study the future of the universe. In the late time the jerk parameter is ended up with $j=1$ for different value of $\alpha$.

\begin{figure}[H]
\centering
\includegraphics[width=8.5 cm]{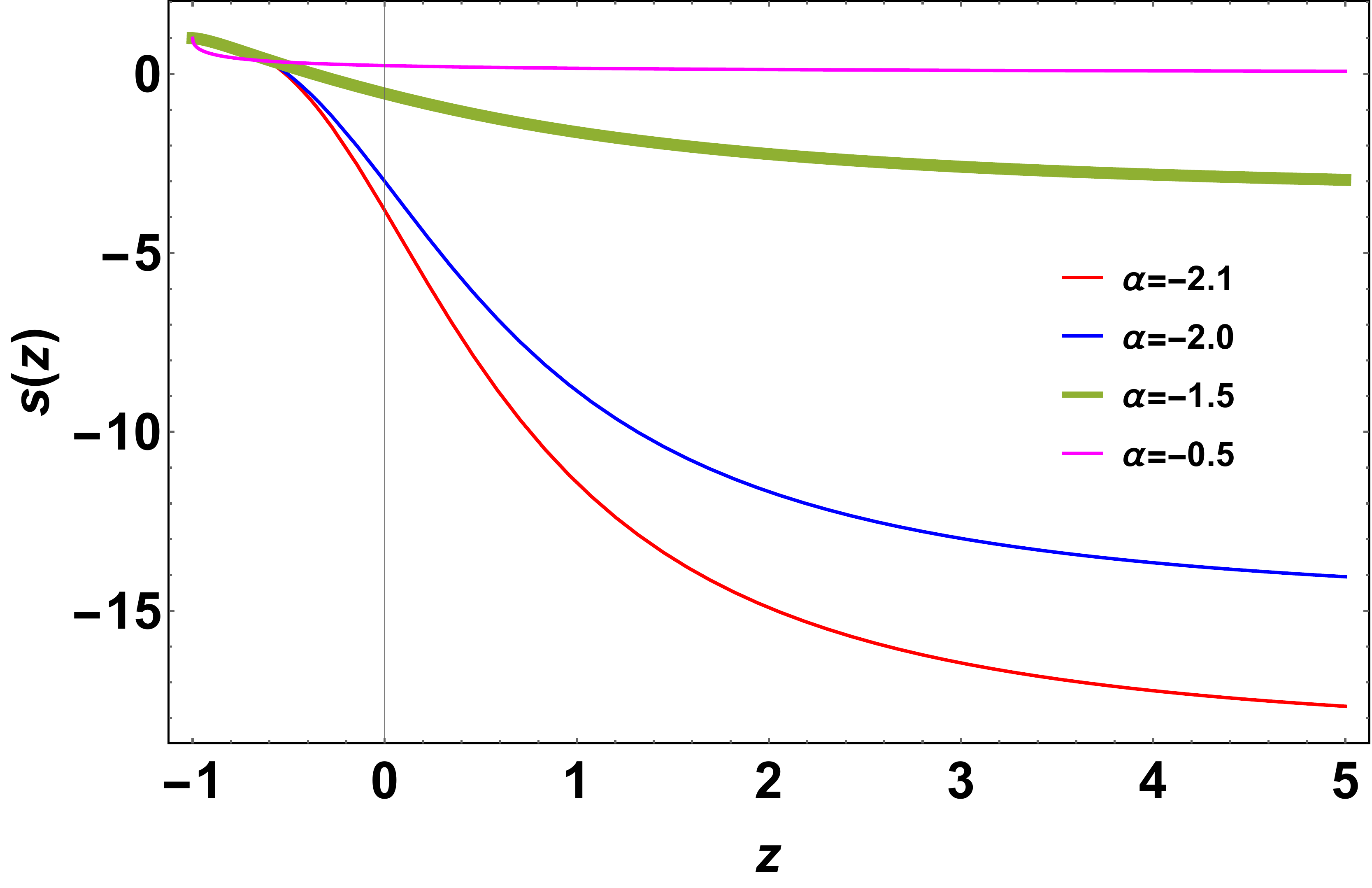}
\caption{Snap parameter as a function of redshift.}
\label{f1b}
\end{figure}
%The evolution of snap parameter $s$ with respect to redshift $z$ shown in Fig.\ref{f1b}. The value of $s$ lies in the negative range during the early stage and then it goes to positive range with the evolution of the universe. Also, it is observed that the transition is delayed and depends on the value of $\alpha$. Fig.\ref{f1c} shows the evolution of lerk parameter $l$ for different values of $\alpha$. It's value takes positive throughout the evolution of the universe. In addition, when $z$ tends to $-1$, $j,k$, and $l$ approaches to $1$.

\section{Cosmology with Teleparallel Gravity}\label{IV}

\subsection{ Hybrid Teleparallel Gravity}

For the first case, we presume the functional form of teleparallel gravity
to be 
\begin{equation}\label{1}
f(T)=e^{mT}T^n,
\end{equation}
where $m\geq0$ and $n$ constants. Interestingly, this model takes power-law
and exponential forms depending on the values of $n$ and $m$. Particularly:

\begin{itemize}
\item For $m=0$ Eq. \eqref{1} reduces to $f(T)=T^n$ (power law).

\item For and $n=0$, Eq. \eqref{1} reduces to $f(T)=e^{mT}$ (exponential).
\end{itemize}

Using Eq. \eqref{1} in Eq. \eqref{4a} and Eq. \eqref{4b}, the expressions of
energy density $\rho$, pressure $p$ and EoS parameter $\omega$ reads
respectively as 
\begin{equation} \label{2}
\rho=3K+6^n(-K)^ne^{-6mK}\left(\frac{1}{2}-n+6nK\right),
\end{equation}
\begin{widetext}
\begin{multline}\label{3}
p=-2\left(\alpha K-\frac{e^{-t\alpha \beta}\alpha\beta^2}{-1+e^{-t\alpha\beta}}\right)\times
\left\lbrace -1+(-6K)^ne^{-6mK}\left[m+4mn-12Km^2-\frac{n}{6K}-\frac{n(n-1)}{3K}\right]\right\rbrace \\
-3K-6^n(-K)^ne^{-6mK}\left(\frac{1}{2}-n+6nK\right)
\end{multline}
\begin{multline}\label{4}
\omega=-1-2\left(\alpha K-\frac{e^{-t\alpha \beta}\alpha\beta^2}{-1+e^{-t\alpha\beta}}\right)\times 
\left\lbrace -1+(-6K)^ne^{-6mK}\left[m+4mn-12Km^2-\frac{n}{6K}-\frac{n(n-1)}{3K}\right]\right\rbrace\\
\times\left\lbrace3K+6^n(-K)^ne^{-6mK}\left(\frac{1}{2}-n+6nK\right)\right\rbrace^{-1}
\end{multline}
\end{widetext}
where $K=\frac{e^{-2t\alpha\beta}\beta^2}{(-1+e^{-t\alpha\beta})^2}$. 
\begin{figure}[H]
\centering
\includegraphics[width=8.5 cm]{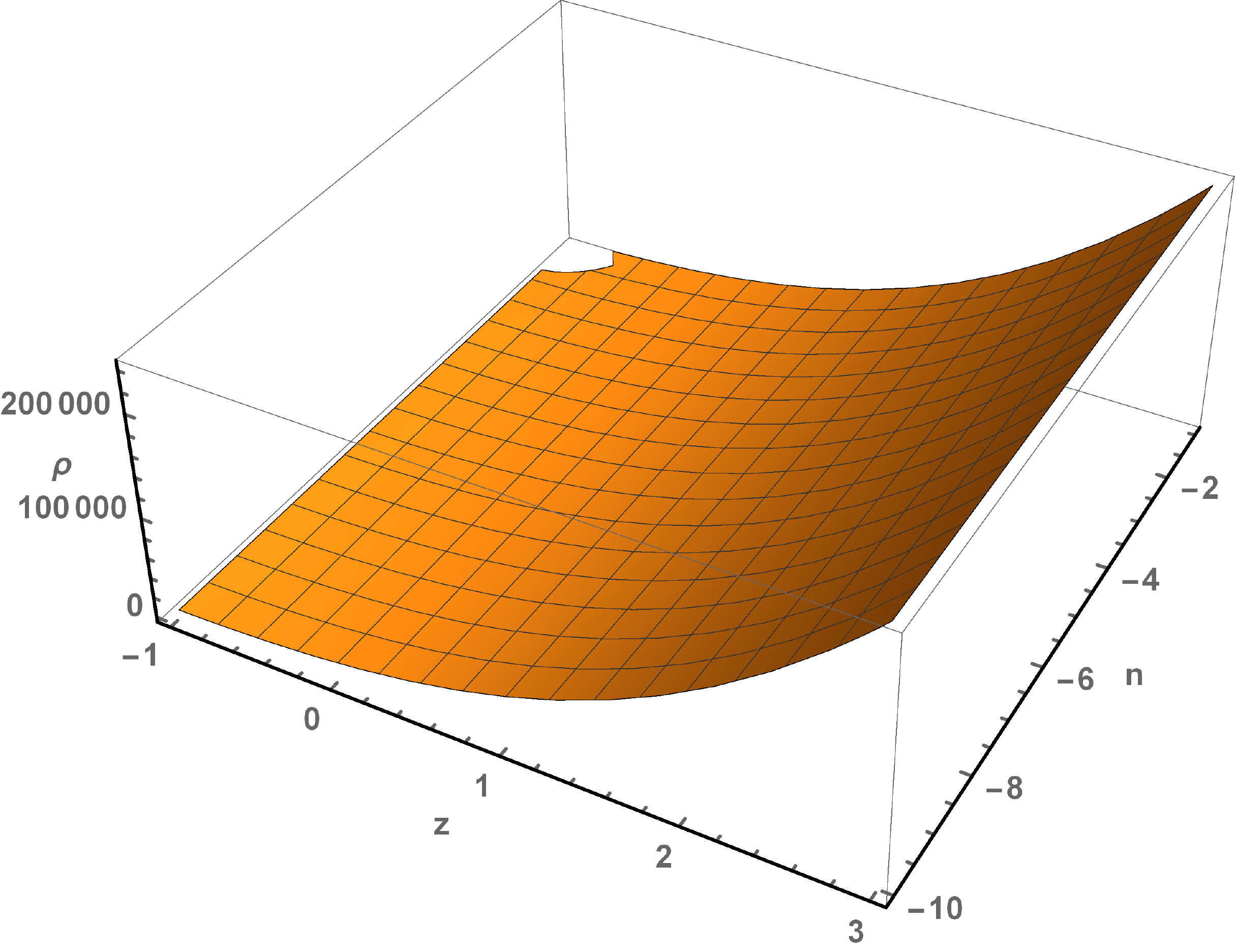}
\caption{Energy density as a function of redshift for $\protect\alpha=-1.5, 
\protect\beta=31.7455 \ \ \& \ \ m=0.00155$.}
\label{f2}
\end{figure}
\begin{figure}[H]
\centering
\includegraphics[width=8.5 cm]{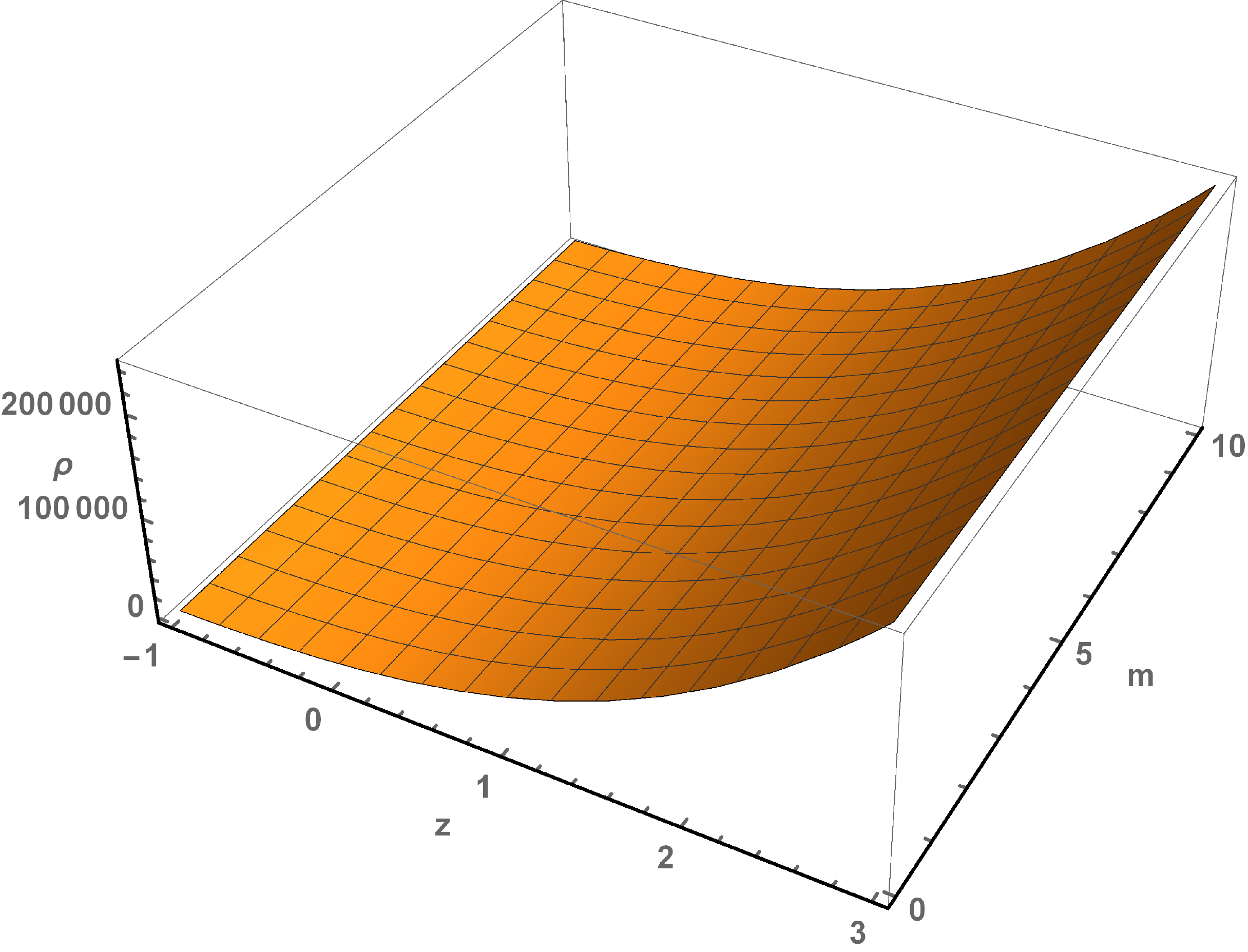}
\caption{Energy density as a function of redshift for $\protect\alpha=-1.5, 
\protect\beta=31.7455 \ \ \& \ \ n=5$.}
\label{f3}
\end{figure}
\begin{figure}[H]
\centering
\includegraphics[width=8.5 cm]{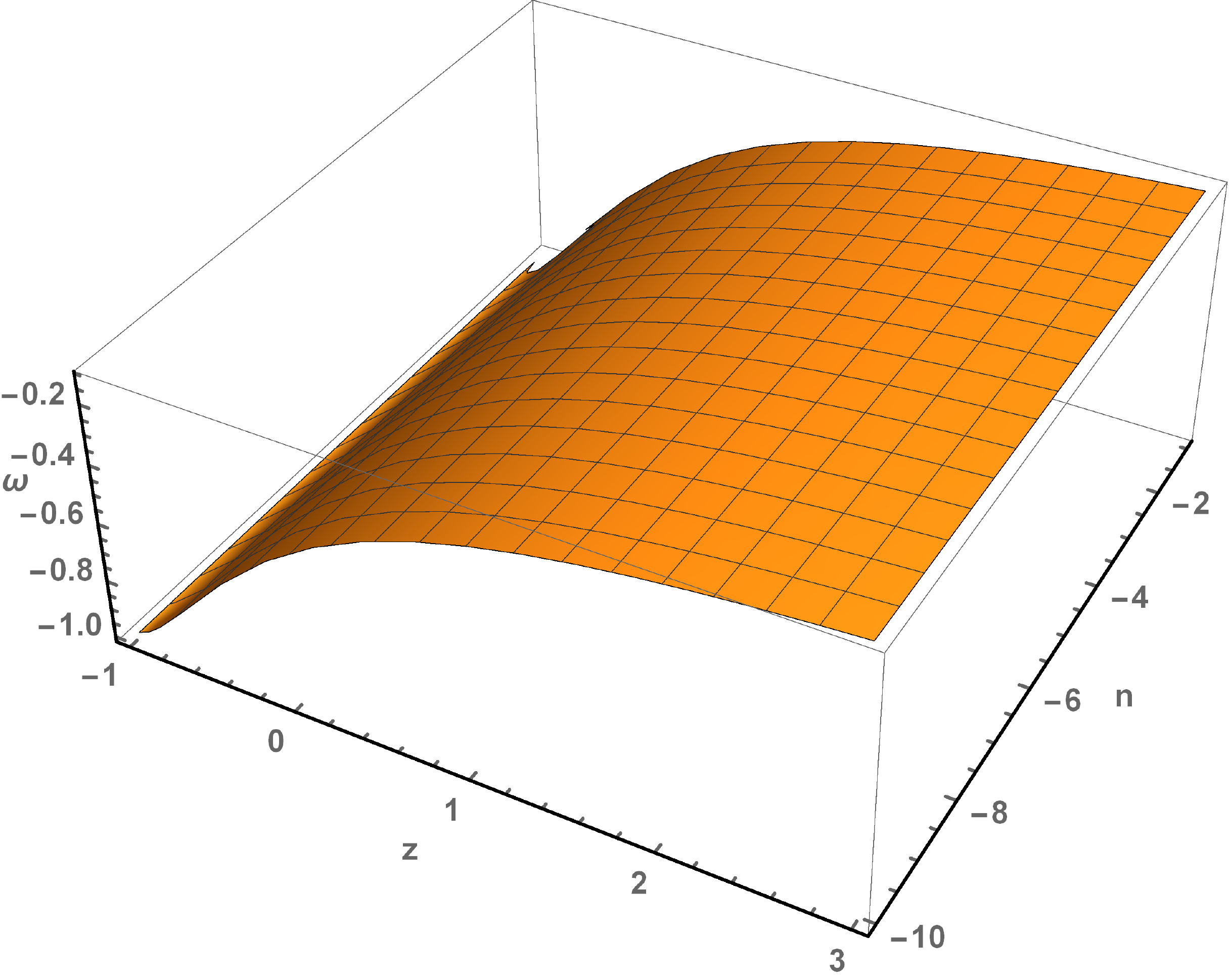}
\caption{EoS parameter as a function of redshift for $\protect\alpha=-1.5, 
\protect\beta=31.7455 \ \ \& \ \ m=0.00155$.}
\label{f4}
\end{figure}

\begin{figure}[H]
\centering
\includegraphics[width=8.5 cm]{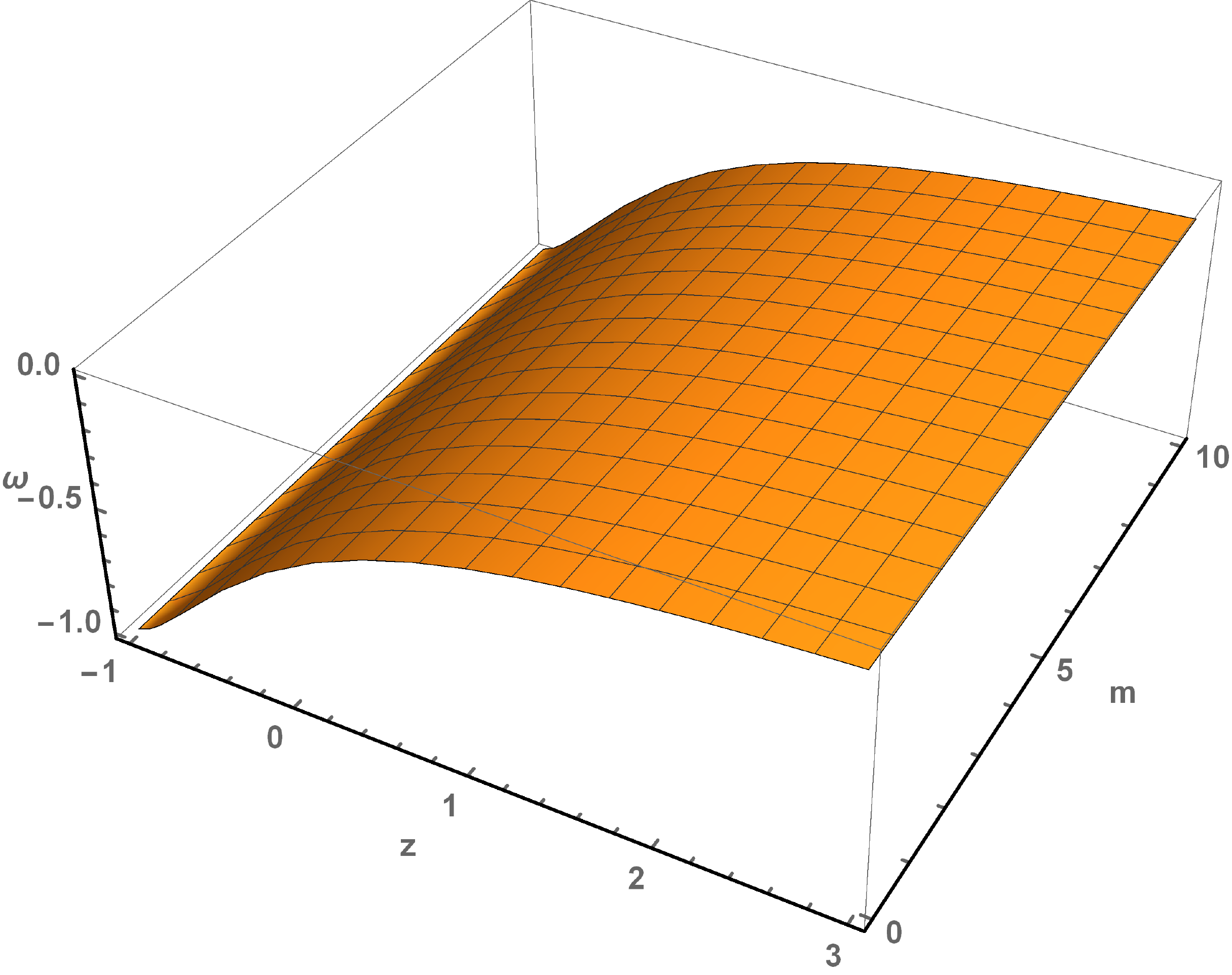}
\caption{EoS parameter as a function of redshift for $\protect\alpha=-1.5, 
\protect\beta=31.7455 \ \ \& \ \ n=5$.}
\label{f5}
\end{figure}

\subsection{ Logarithmic Teleparallel Gravity}

For the second case, we presume the functional form of teleparallel gravity
to be 
\begin{equation}  \label{5}
f(T)=D\log(bT),
\end{equation}
where $D$ and $b<0$ are constants.\newline
Using Eq. \eqref{5} in Eq. \eqref{4a} and Eq. \eqref{4b}, the expressions of
energy density $\rho$, pressure $p$ and EoS parameter $\omega$ reads
respectively as 
\begin{equation}  \label{6}
\rho=-D+3K+\frac{D}{2}\log(6bK),
\end{equation}
\begin{widetext}
\begin{equation}\label{7}
p=D-3K-2\left(1+\frac{D}{6K}\right)\left(\alpha K-\frac{e^{-t\alpha\beta}\alpha\beta^2}{-1+e^{-t\alpha\beta}}\right)-\frac{D}{2}\log(6bK),
\end{equation}
\begin{equation}\label{8}
\omega=-1-2\left(1+\frac{D}{6K}\right)\left(\alpha K-\frac{e^{-t\alpha\beta}\alpha\beta^2}{-1+e^{-t\alpha\beta}}\right)
\times \left\lbrace -D+3K+\frac{D}{2}\log(6bK)\right\rbrace^{-1}
\end{equation}
\end{widetext}

\begin{figure}[H]
\centering
\includegraphics[width=8.5 cm]{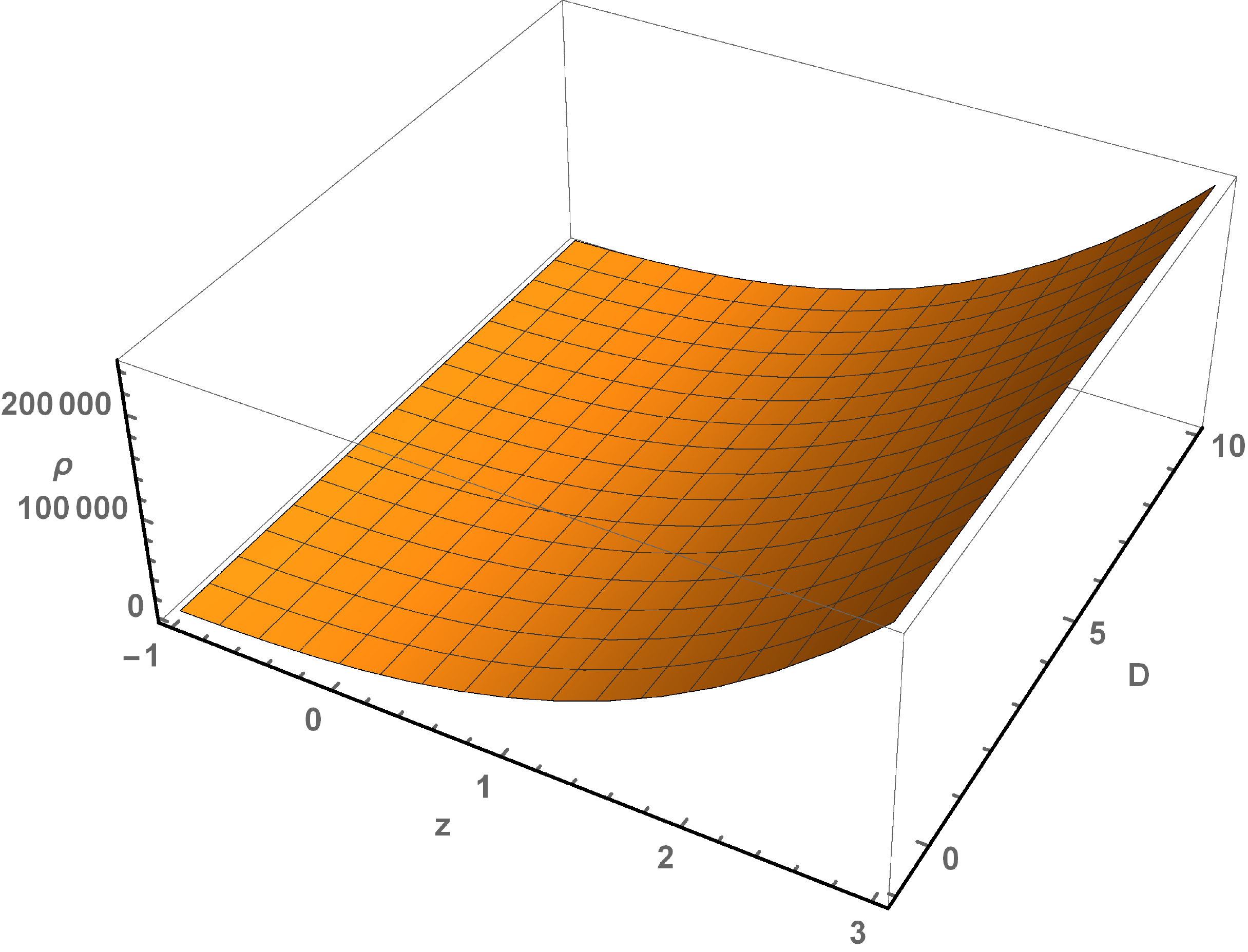}
\caption{Energy density as a function of redshift for $\protect\alpha=-1.5, 
\protect\beta=31.7455 \ \ \& \ \ b=-2$.}
\label{f6}
\end{figure}
\begin{figure}[H]
\centering
\includegraphics[width=8.5 cm]{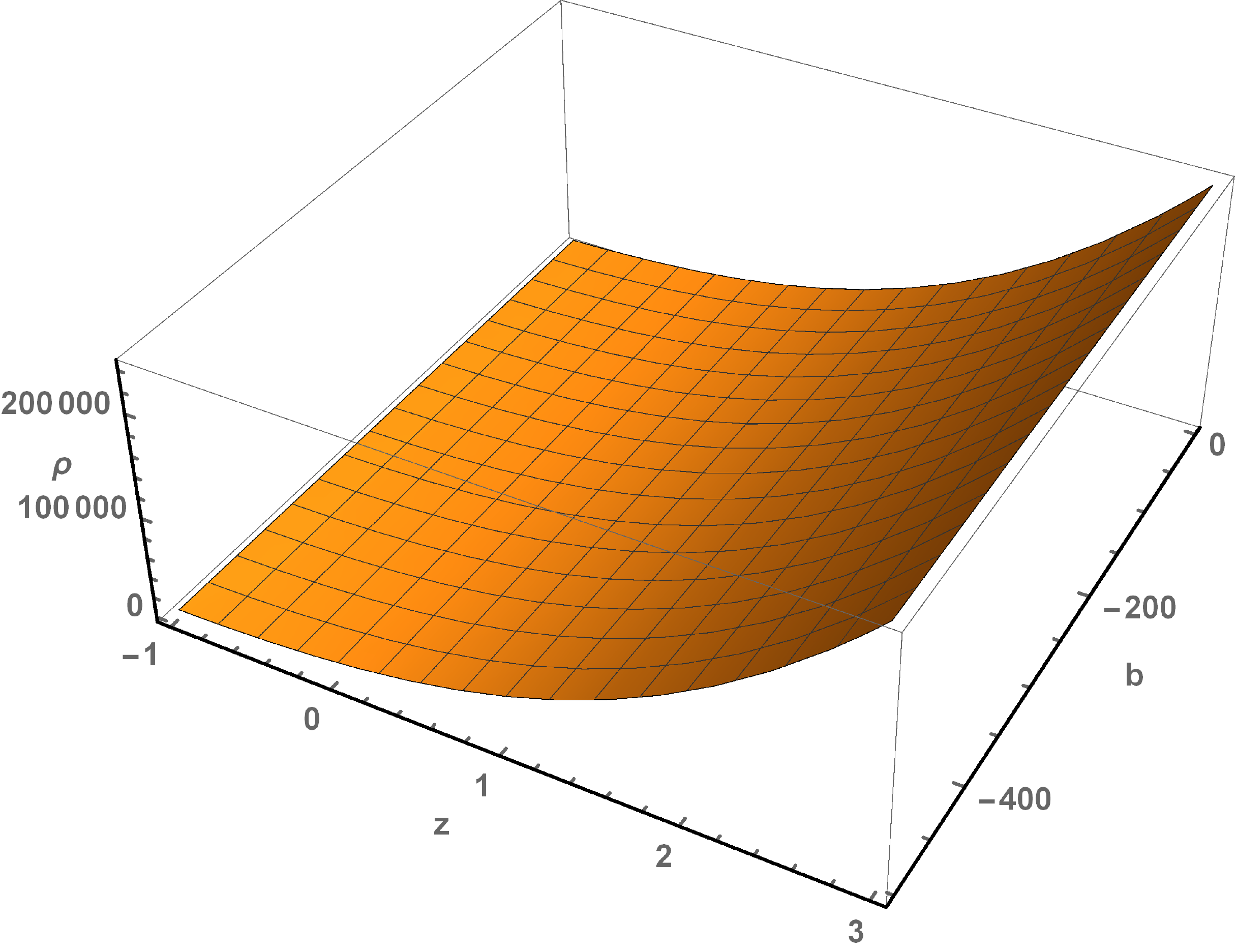}
\caption{Energy density as a function of redshift for $\protect\alpha=-1.5, 
\protect\beta=31.7455 \ \ \& \ \ D=0.2$.}
\label{f7}
\end{figure}
\begin{figure}[H]
\centering
\includegraphics[width=8.5 cm]{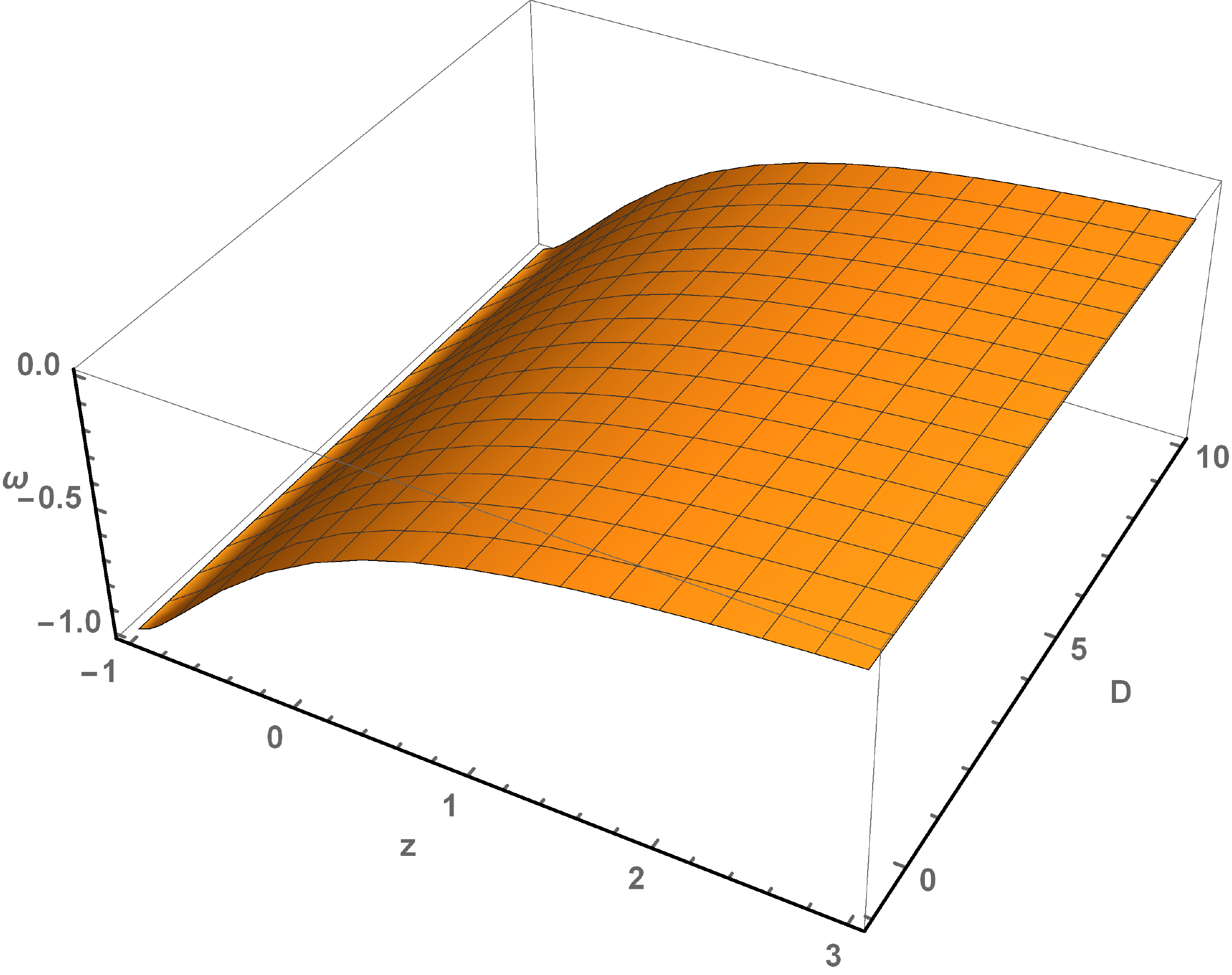}
\caption{EoS parameter as a function of redshift for $\protect\alpha=-1.5, 
\protect\beta=31.7455 \ \ \& \ \ b=-2$.}
\label{f8}
\end{figure}
\begin{figure}[H]
\centering
\includegraphics[width=8.5 cm]{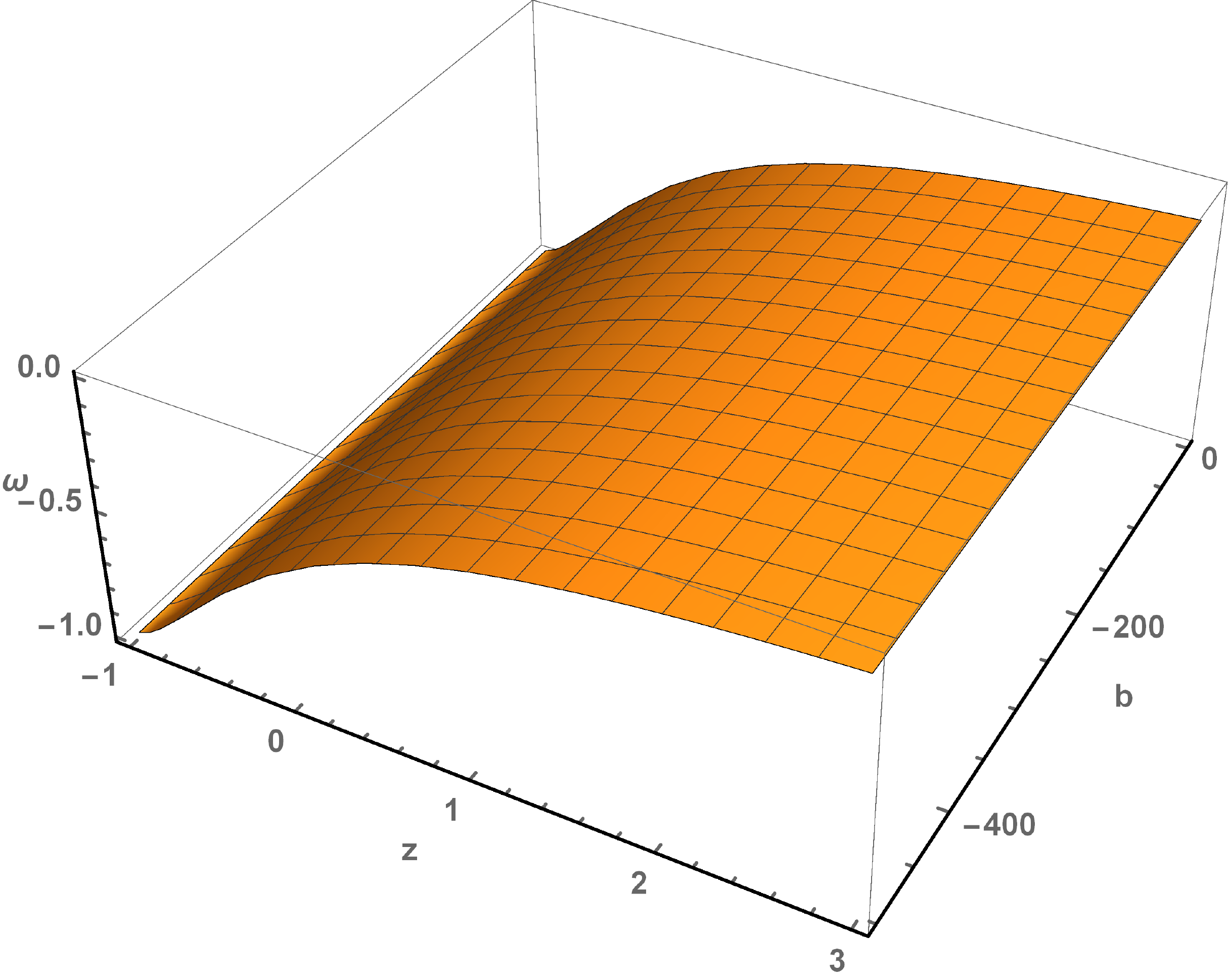}
\caption{EoS parameter as a function of redshift for $\protect\alpha=-1.5, 
\protect\beta=31.7455 \ \ \& \ \ D=0.2$.}
\label{f9}
\end{figure}

\section{Geometrical Diagnostics}\label{V}

\subsection{Statefinder Diagnostics}

Due to the fact that the number of dark energy models are quite large and
increasing on a daily basis, it becomes absolutely necessary to find a
method to distinguish a particular model from the well established DE models
like the $\Lambda$CDM, SCDM, HDE, CG and Quintessence. With that reasoning, 
\cite{sahni} proposed the $\{r,s\}$ diagnostics where $r$ and $s$ are
defined as

\begin{align*}
r=\frac{\dot{\ddot{a}}}{aH^3},
\end{align*}
\begin{align*}
s=\frac{r-1}{3\left(q-\frac{1}{2}\right)},\left(q\neq\frac{1}{2}\right).
\end{align*}
Different combinations of $r$ and $s$ represent different dark energy
models. Particularly:

\begin{itemize}
\item For $\Lambda$CDM$\rightarrow$ $(r=1,s=0)$.

\item For SCDM$\rightarrow$ $(r=1,s=1)$.

\item For HDE$\rightarrow$ $(r=1,s=\frac{2}{3})$.

\item For CG$\rightarrow$ $(r>1,s<0)$.

\item For Quintessence $\rightarrow$ $(r<1,s>0)$.
\end{itemize}

The idea behind $\{r,s\}$ diagnostics tool is that different dark energy
models exhibit different trajectories in the $\{r,s\}$ plane. The deviation
from the point $\{r,s\} = \{0,1\}$ represent deviation from the well agreed $%
\Lambda$CDM model. Furthermore, the values of $r$ and $s$ could in principle
be inferred from observations \cite{sahni17to18} and therefore could be very
useful in discriminating dark energy models in the near future.\newline
The expression of $r$ and $s$ parameters for our model reads 
\begin{equation}
r=1+\frac{\alpha\left(\frac{1}{1+z}\right)^{\alpha}\left\lbrace
3+\alpha+\left(\frac{1}{1+z}\right)^{\alpha}(3+2\alpha)\right\rbrace}{%
\left\lbrace 1+\left(\frac{1}{1+z}\right)^{\alpha}\right\rbrace^2}
\end{equation}
\begin{equation}
s=\frac{\alpha}{3}\left\lbrace -2+\frac{1}{1+\left(\frac{1}{1+z}%
\right)^{\alpha}}+\frac{3}{3+\left(\frac{1}{1+z}\right)^{\alpha}(3+2\alpha)}%
\right\rbrace
\end{equation}

\begin{figure}[H]
\centering
\includegraphics[width=8.5 cm]{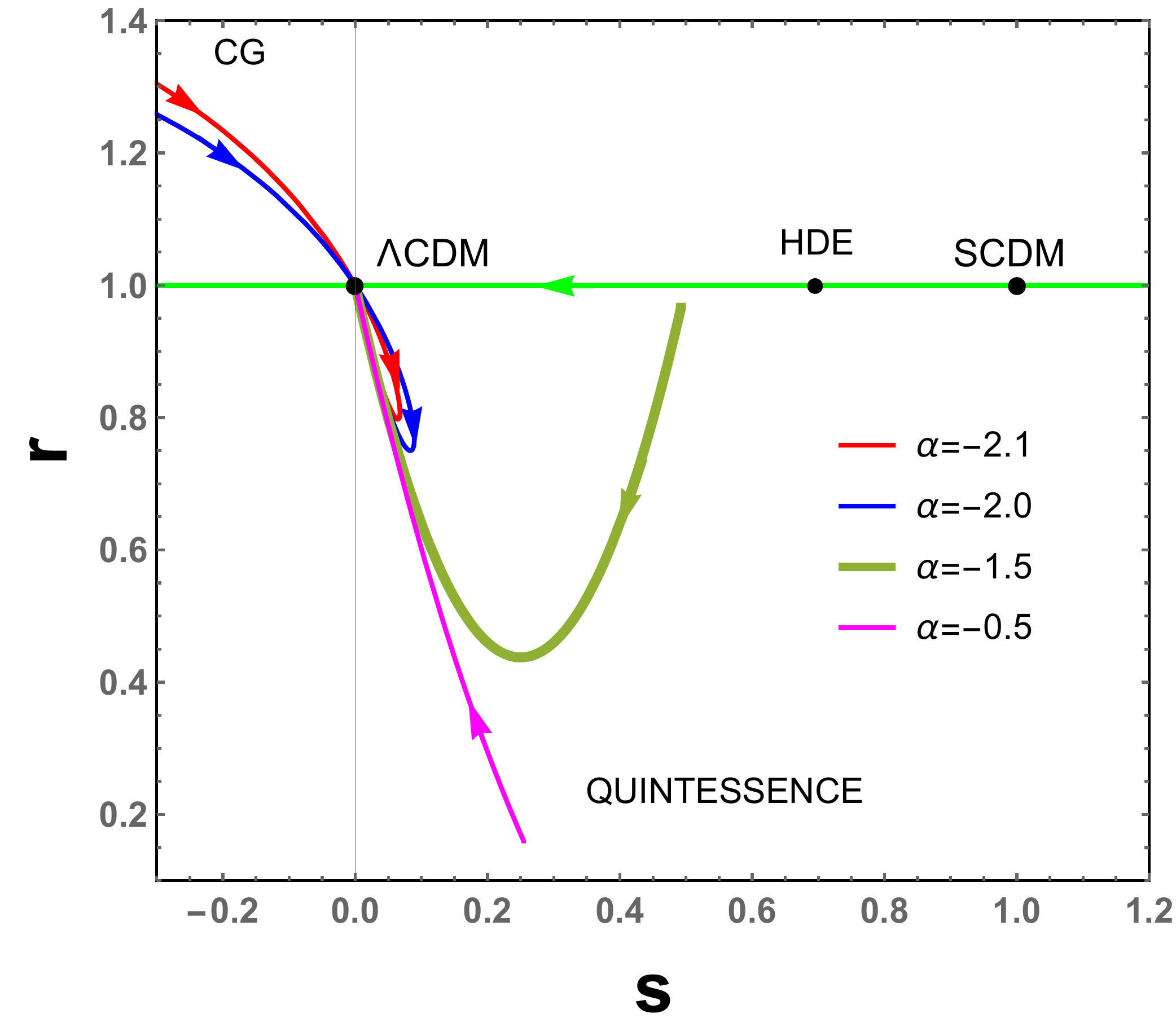}
\caption{$\{r,s\}$ plane for the redshift range $z[-1,5]$ for different
values of $\protect\alpha$.}
\label{f12}
\end{figure}

In Fig. \ref{f12}, the $\{r,s\}$ plane is shown for the parametrization %
\eqref{5e} where the arrows indicate the direction of temporal evolution.
The model is observed to deviate significantly from the point $(0,1)$
initially and is extremely sensitive to the value of $\alpha$. For $%
\alpha\leq-2$, the model initially starts its journey from the territory of
Chaplygin gas $(r>1,s<0)$ and approaches towards $\Lambda$CDM at late times.
For $\alpha>-1$, the model at high redshifts stays in the Quintessence
region but again approaches towards $\Lambda$CDM. Interestingly, for $%
\alpha=-1.5\sim-1.5$ which is the chi-square value, we observed at high
redshifts, the model to be very close to the point $(r=1,s=\frac{2}{3})$
which is the region of HDE. However, at late times the model is observed to
coincide with ($r=1,s=0$). Therefore, the parametrization used in this work
is interesting and warrants further attention. 
In addition to the $\{r,s\}$ plane, we construct the $\{r,q\}$ plane to get
additional understanding of the parametrization \eqref{5e}. In $\{r,q\}$
plane, the solid line in the middle represents the evolution of the $\Lambda$%
CDM universe and also divide the plane into two sections. The upper section
belong to Chaplygin gas models and the lower section to Quintessence models. 
\begin{figure}[H]
\centering
\includegraphics[width=8.5 cm]{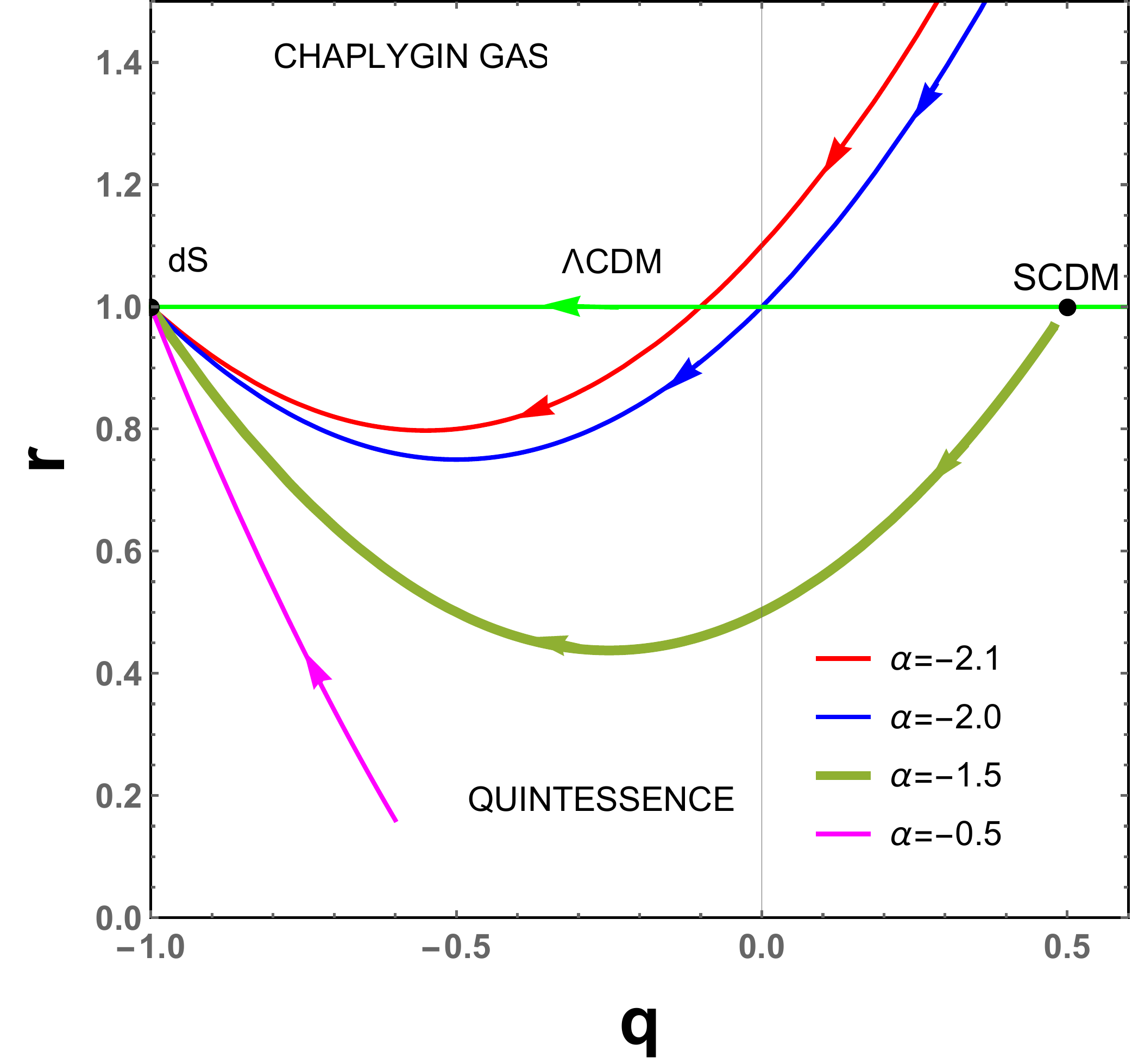}
\caption{$\{r,q\}$ plane for the redshift range $z[-1,5]$ for different
values of $\protect\alpha$.}
\label{frq}
\end{figure}
We observe from Fig. \ref{frq}, that except for $\alpha=-0.5$, all the
profiles starts from $r>1,q>0$ which is very close to SCDM universe,
followed by the region $r<1,-1<q<0$ and finally approaches the de-Sitter
expansion with $r=1,q=-1$. However, for $\alpha<-1$, $q$ is always negative
and therefore the profile does not start from the SCDM universe.

\subsection{Om Diagnostic}

\begin{figure}[H]
\centering
\includegraphics[width=8.5 cm]{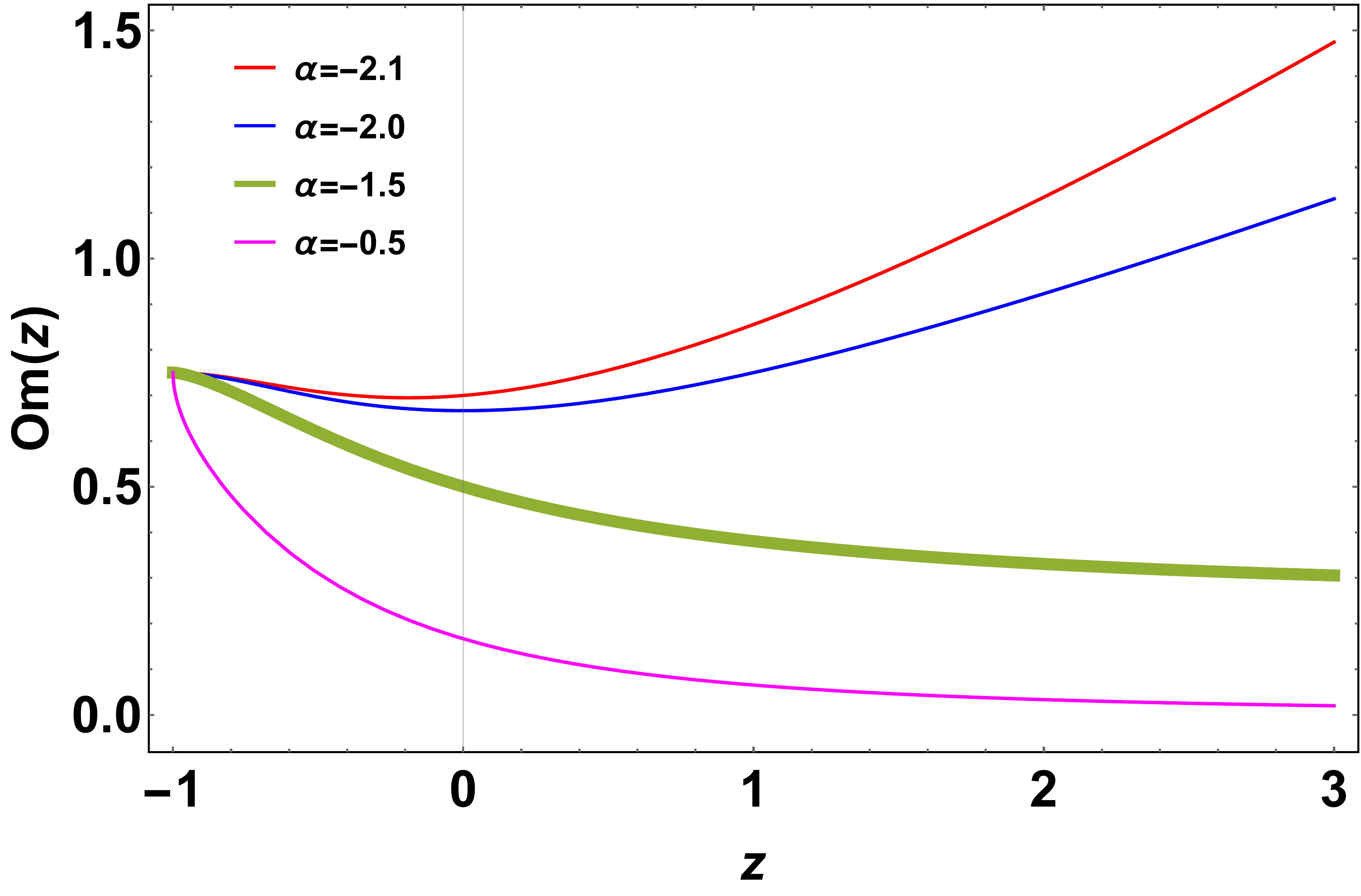}
\caption{ $Om(z)$ for different values of $\protect\alpha$.}
\label{f13}
\end{figure}

Another very useful diagnostic tool constructed from the Hubble parameter is
the $Om$ diagnostic which essentially provide a null test of the $\Lambda$%
CDM model \cite{Omsahni}. This tool easily captures the dynamical nature of
dark energy models from the slope of $Om(z)$. If the slope of this
diagnostic tool were to be positive, it would imply a Quintessence nature ($%
\omega>-1$) whereas the opposite would prefer a Phantom nature ($\omega<-1$%
). Interestingly, only when the nature of the dark energy model coincide
with that of the cosmological constant ($\omega=-1$), the slope is constant
with respect to redshift. It is defined as

\begin{equation}
Om(z)=\frac{\left(\frac{H(z)}{H_0}\right)^2-1}{z^3+3z^2+3z}
\end{equation}
From Fig. \ref{f13}, we observe a negative slope for $\alpha>-2$ and
therefore represents a dark energy model which is Phantom in nature.
Nonetheless, for $\alpha\leq-2$, $Om(z)$ increases with redshift and
therefore represents an Quintessence dark energy model. Hence, the value of $%
\alpha$ dictates the nature of the underlying dark energy model represented
by the parametrization \eqref{5e}.

\section{Energy Conditions}\label{VI}

Based upon the Raychaudhuri equation, the energy conditions are essential to
describe the behavior of the compatibility of timelike, lightlike or
spacelike curves \cite{sahoo} and often used to understand the dreadful
singularities \cite{non39}. Energy conditions in teleparallel gravity have
been studied in \cite{energy}. Energy conditions also provide the corners in
parameter spaces since they violate, for instance, in presence of
singularities. They are defined as:

\begin{itemize}
\item Strong energy conditions (SEC): Gravity is always attractive and
therefore $\rho+3p\geq 0$;

\item Weak energy conditions (WEC): Energy density should always be
positive, i.e., $\rho\geq 0, \rho+p\geq 0$;

\item Null energy condition (NEC): Minimum requirement for the fulfilment of
SEC and WEC, i.e., $\rho+p\geq 0$;

\item Dominant energy conditions (DCE): Energy density is always positive
and independent of the observer's reference frame, i.e., $\rho\geq 0,
|p|\leq \rho$.
\end{itemize}

\begin{figure}[H]
\centering
\includegraphics[width=8.5 cm]{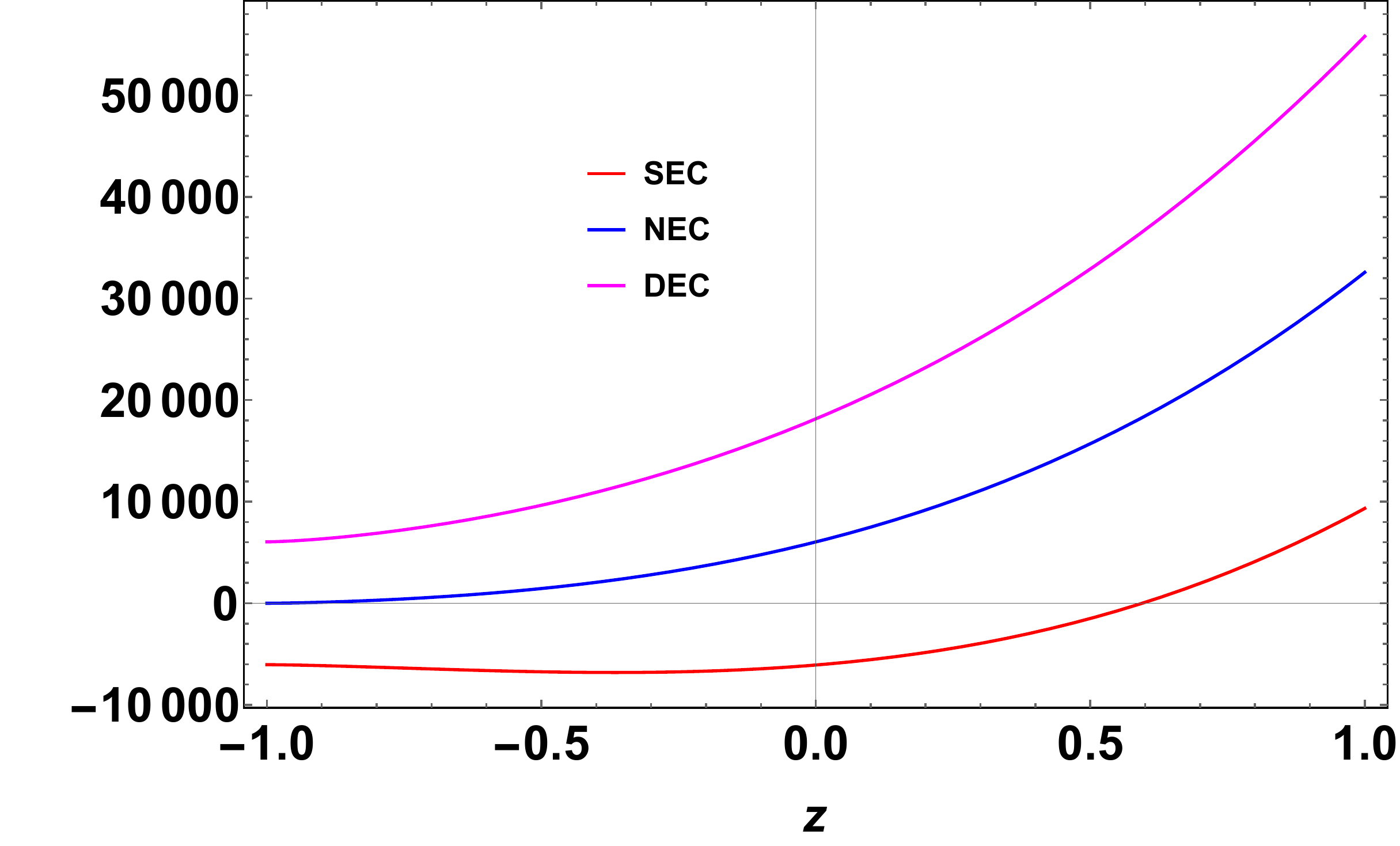}
\caption{ ECs as a function redshift $z$ for $\protect\beta=31.7455,\protect%
\alpha=-1.5$, $m=0.00155$ \& $n=-5$ for hybrid teleparallel gravity.}
\label{f14}
\end{figure}

\begin{figure}[H]
\centering
\includegraphics[width=8.5 cm]{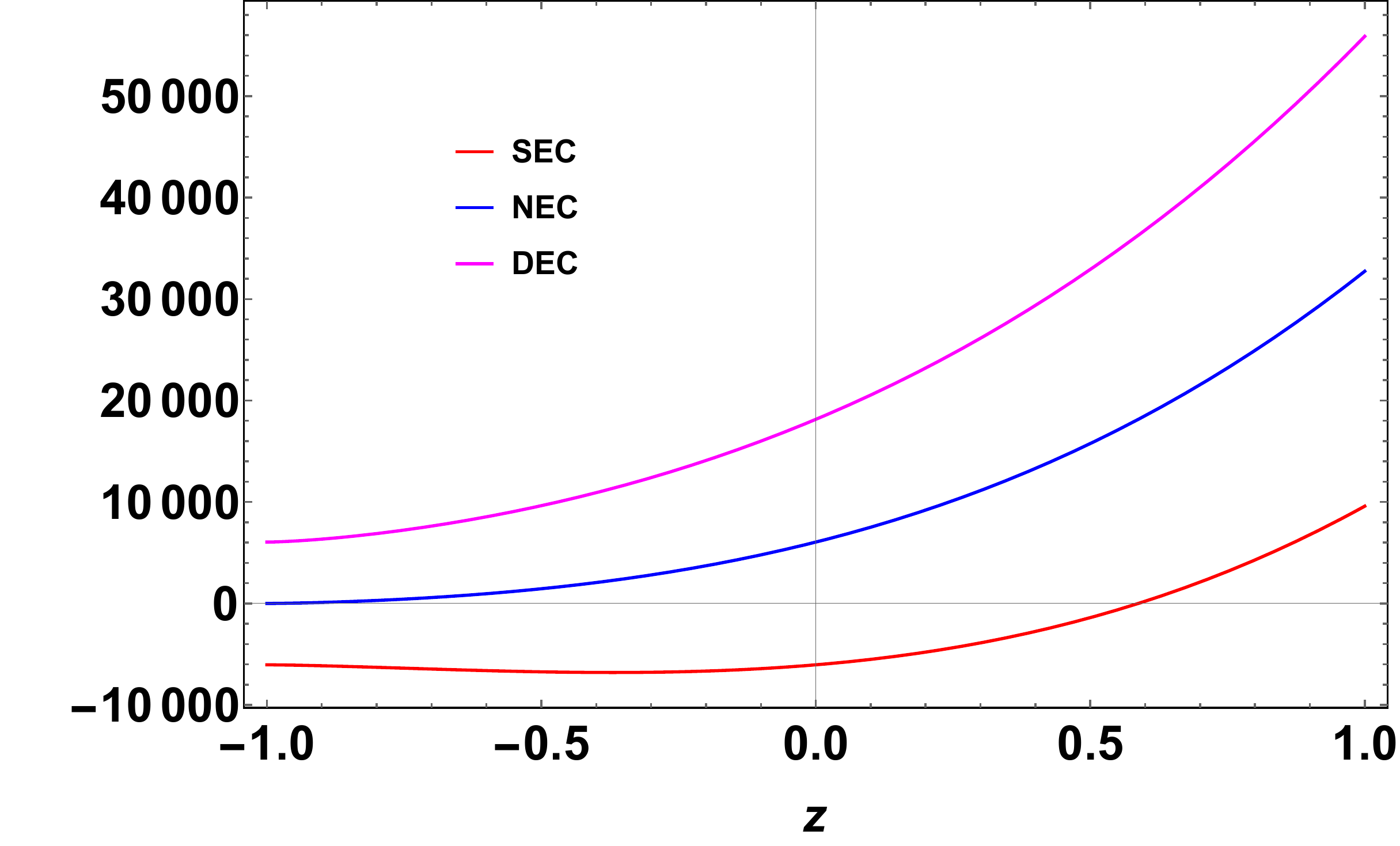}
\caption{ECs as a function redshift $z$ for $\protect\beta=31.7455,\protect%
\alpha=-1.5, b=-2$ \& $d=0.2$ for logarithmic teleparallel gravity.}
\label{f15}
\end{figure}

Energy conditions for both the teleparallel gravity models are presented in
Fig. \ref{f14}-\ref{f15}.


\section{Observational Constraints}\label{VII}

In order to find the best fit value of the model parameters of our obtained
models, we need to constrain the parameters with some available datasets.
Here, we use three datasets, namely, Hubble datasets with $57$ datapoints,
Supernovae datasets consisting of $580$ data points from Union$2.1$
compilation datasets and Baryonic Aucostic Oscillation (BAO) datasets. We
use the Bayesian statistics for our analysis.

\subsection{Hubble parameter H(z)}

Recently, Sharov and Vasiliev \cite{sharov} compiled a list of $57$ points
of measurements of the Hubble parameter at in the redshift range $%
0.07\leqslant z\leqslant 2.42$, measured by extraction of $H(z)$ from
line-of-sight BAO data including the analysis of correlation functions of
luminous red galaxies \cite{Hz1} and $H(z)$ estimations from differential
ages $\vartriangle t$ of galaxies (DA method) \cite{Hz2}. (See the Appendix
in \cite{sharov} for full list of tabulated datasets). Chi square test is
used to constrain the model parameters parameters given by

\begin{equation}
\chi _{OHD}^{2}(p_{s})=\sum_{i=1}^{57}\frac{%
[H_{th}(p_{s},z_{i})-H_{obs}(z_{i})]^{2}}{\sigma _{H(z_{i})}^{2}}
\label{chi}
\end{equation}%
where $H_{th}(p_{s},z_{i})$ denotes the Hubble parameter at redshift $z_{i}$
predicted by the models with $p_{s}$ denoting the parameter space ($\alpha $
here in our model), $H_{obs}(z_{i})$ is the $i$-th measured one and $\sigma
_{H(z_{i})}$ is its uncertainty. We also take a prior as $H_{0}=67.8$ (Plank
result predicted value) for our analysis.

\subsection{Type Ia Supernova}

Further we consider, the $580$ points of Union$2.1$ compilation supernovae
datasets \cite{SNeIa} for our analysis for which the chi square formula is
given as,

\begin{equation}
\chi _{SN}^{2}(\mu _{0},p_{s})=\sum\limits_{i=1}^{580}\frac{[\mu _{th}(\mu
_{0},p_{s},z_{i})-\mu _{obs}(z_{i})]^{2}}{\sigma _{\mu (z_{i})}^{2}},
\label{chisn}
\end{equation}%
where $\mu _{th}$ and $\mu _{obs}$ are correspondingly the theoretical and
observed distance modulus for the model and the standard error is $\sigma
_{\mu (z_{i})}$. The distance modulus $\mu (z)$ is defined to be $\mu
(z)=m-M=5LogD_{l}(z)+\mu _{0},$where $m$ and $M$ are the apparent and
absolute magnitudes of any standard candel (supernovae of type \textit{Ia}
here) respectively. Luminosity distance $D_{l}(z)$ and the nuisance
parameter $\mu _{0}$ are given by $D_{l}(z)=(1+z)H_{0}\int_{0}^{z}\frac{1}{%
H(z^{\ast })}dz^{\ast }$ and $\mu _{0}=5Log\Big(\frac{H_{0}^{-1}}{Mpc}\Big)%
+25$ respectively. In order to calculate luminosity distance, we have
restricted the series of $H(z)$ up to tenth term only and then integrated
the approximate series to obtain the luminosity distance.

\subsection{Baryon Acoustic Oscillations}

Finally, we consider a sample of BAO distances measurements from surveys of
SDSS(R) \cite{padn}, 6dF Galaxy survey \cite{6df}, BOSS CMASS \cite{boss}
and three parallel measurements from WiggleZ \cite{wig}. In the context of
BAO measurements, the distance redshift ratio $d_{z}$ is given as, 
\begin{equation}
d_{z}=\frac{r_{s}(z_{\ast })}{D_{v}(z)},  \label{drr}
\end{equation}%
where $r_{s}(z_{\ast })$ is the co-moving sound horizon at the time photons
decouple and $z_{\ast }$ indicates the photons decoupling redshift i.e. $%
z_{\ast }=1090$ \cite{adep}. Moreover, $r_{s}(z_{\ast })$ is assumed same as
considered in the reference \cite{waga} together with the dilation scale is
given by $D_{v}(z)=\big(\frac{d_{A}^{2}(z)z}{H(z)}\big)^{\frac{1}{3}}$,
where $d_{A}(z)$ is the angular diameter distance. The $\chi _{BAO}^{2}$
values corresponding to BAO measurements are discussed in details in \cite%
{gio} and the chi square formula is given by,%
\begin{equation}
\chi _{BAO}^{2}=A^{T}C^{-1}A,  \label{OBAO}
\end{equation}%
where the matrix $A$ is given\ by 
\[
A=\left[ {%
\begin{array}{cc}
\frac{d_{A}(z_{\ast })}{D_{v}(0.106)}-30.84 &  \\ 
\frac{d_{A}(z_{\ast })}{D_{v}(0.35)}-10.33 &  \\ 
\frac{d_{A}(z_{\ast })}{D_{v}(0.57)}-6.72 &  \\ 
\frac{d_{A}(z_{\ast })}{D_{v}(0.44)}-8.41 &  \\ 
\frac{d_{A}(z_{\ast })}{D_{v}(0.6)}-6.66 &  \\ 
\frac{d_{A}(z_{\ast })}{D_{v}(0.73)}-5.43 &  \\ 
& 
\end{array}%
}\right] ,
\]%
and $C^{-1}$ representing the inverse of covariance matrix $C$ given as in
the reference \cite{gio} adopting the correlation coefficients presented in 
\cite{hing} as 
\[
C^{-1}=\left[ {%
\begin{array}{cccccc}
0.52552 & -0.03548 & -0.07733 & -0.00167 & -0.00532 & -0.00590 \\ 
-0.03548 & 24.97066 & -1.25461 & -0.02704 & -0.08633 & -0.09579 \\ 
-0.07733 & -1.25461 & 82.92948 & -0.05895 & -0.18819 & -0.20881 \\ 
-0.00167 & -0.02704 & -0.05895 & 2.91150 & -2.98873 & 1.43206 \\ 
-0.00532 & -0.08633 & -0.18819 & -2.98873 & 15.96834 & -7.70636 \\ 
-0.00590 & -0.09579 & -0.20881 & 1.43206 & -7.70636 & 15.28135 \\ 
&  &  &  &  & 
\end{array}%
}\right] .
\]

Below, we have shown a comparision of our obtained model with the $\Lambda $%
CDM model together with error bars due to the $57$ points of $H(z)$ datasets
and the $580$ points of Union$2.1$ compilation datasets.

\begin{figure}[H]
\label{Error-sn-teleparallel}
\par
\begin{center}
$%
\begin{array}{c@{\hspace{.1in}}c}
\includegraphics[width=3.0 in, height=2.5 in]{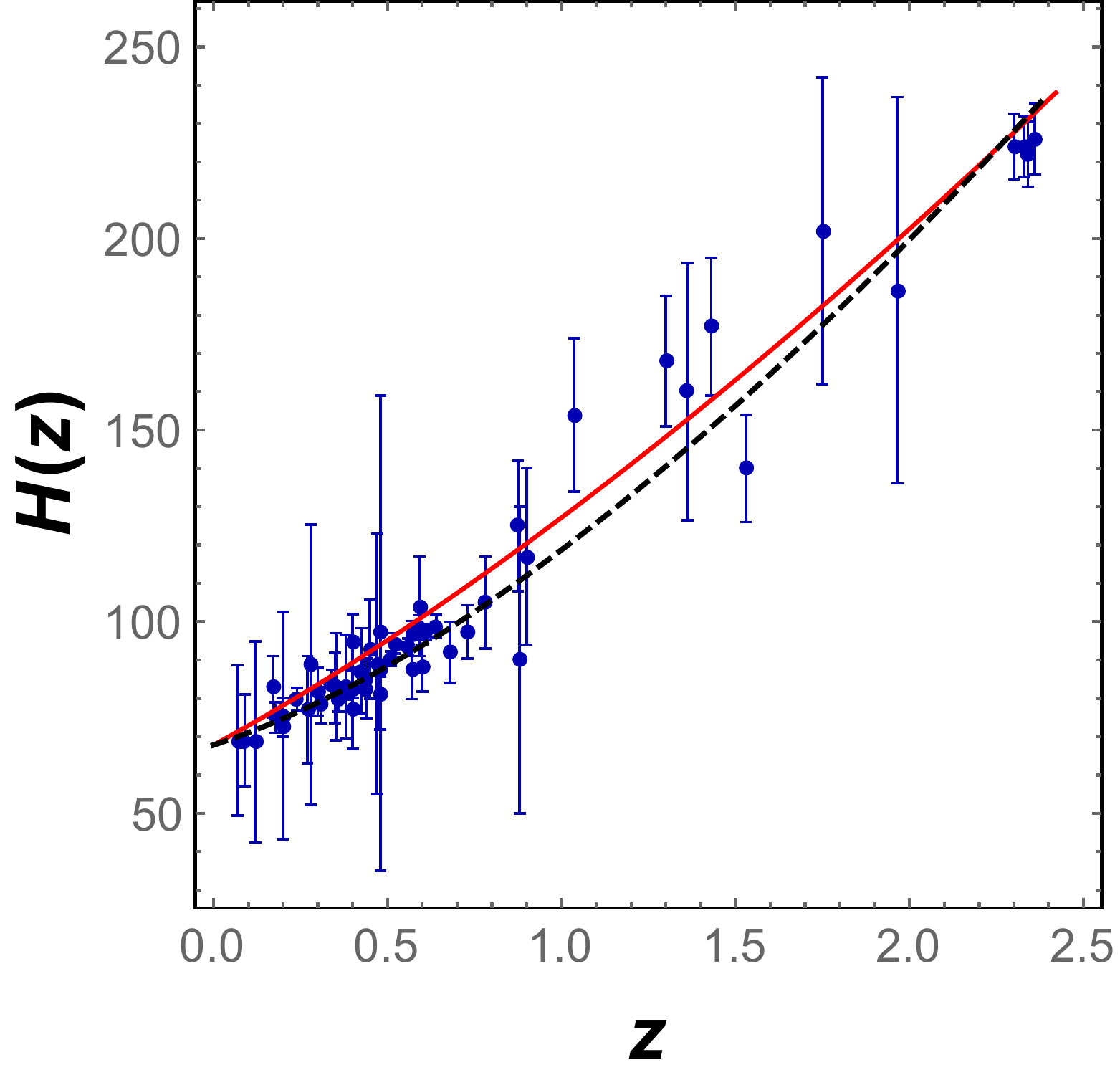} & %
\includegraphics[width=3.0 in, height=2.5 in]{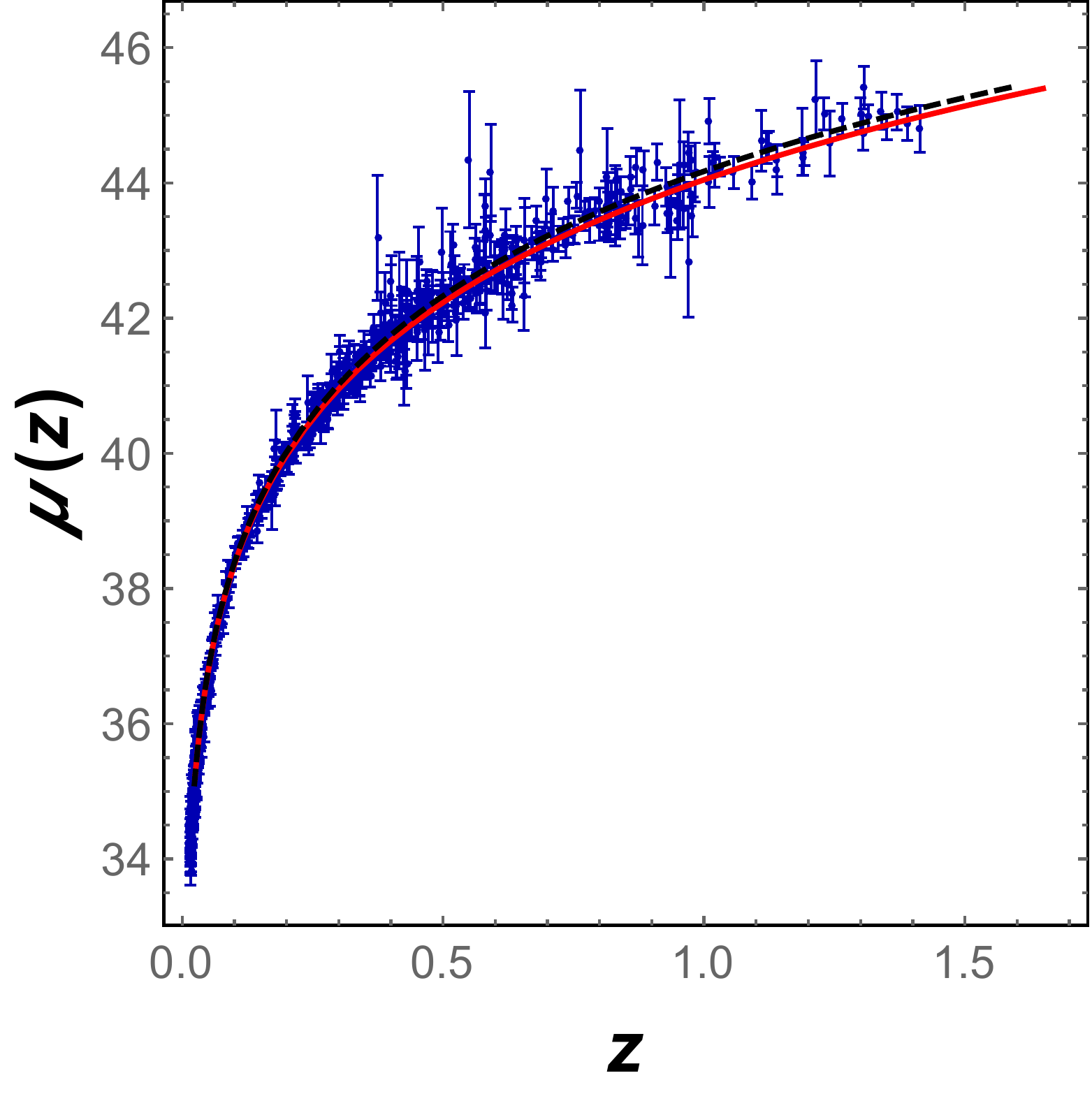} \\ 
\mbox (a) & \mbox (b)%
\end{array}%
$%
\end{center}
\caption{ Figures (a) and (b) are respectively the error bar plots of $57$
points of $H(z)$ datasets and $580$ points of Union$2.1$ compilation
supernovae datasets together with our obtained model (solid red lines) and $%
\Lambda CDM$ model (black dashed lies).}
\end{figure}

Next, we have shown the likelihood contours for the model parameter $\alpha $
and Hubble constant $H_{0}$ with errors at $1$-$\sigma $, $2$-$\sigma $ and $%
3$-$\sigma $ levels in the $\alpha $-$H_{0}$ plane. The best fit constrained
values of $\alpha $ and $H_{0}$ are found to be $\alpha =-1.497294$ \& $%
H_{0}=63.490604$ due to $H(z)$ datasets only with $\chi _{\min
}^{2}=31.333785$ and $\alpha =-1.503260$ \& $H_{0}=63.361612$ due to joint
datasets $H(z)$ + $SNeIa$ + $BAO$ with $\chi _{\min }^{2}=650.312968$
respectively.

\begin{figure}[H]
\begin{center}
$%
\begin{array}{c@{\hspace{.1in}}c}
\includegraphics[width=3.0 in, height=2.5 in]{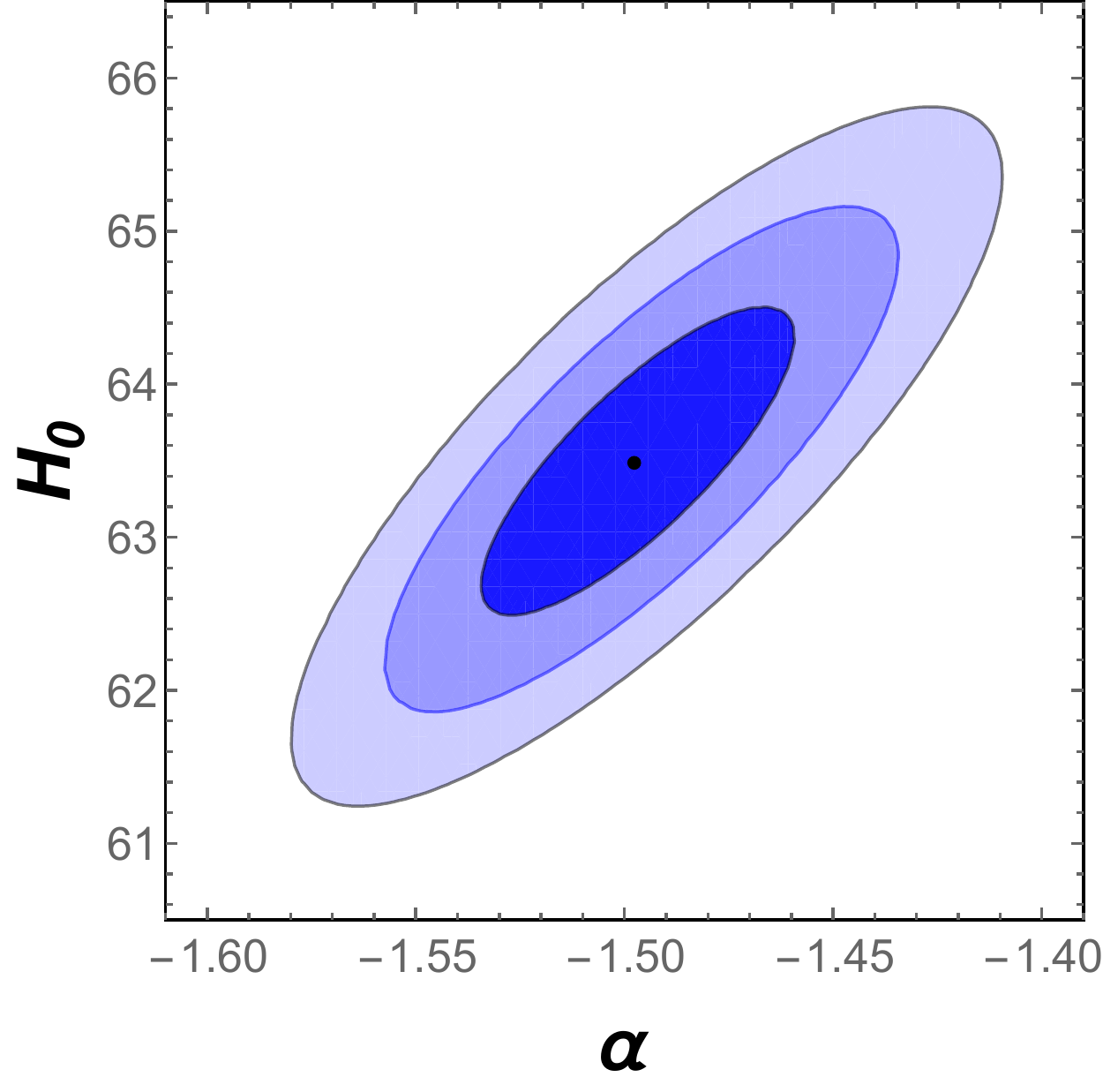} & %
\includegraphics[width=3.0 in, height=2.5
in]{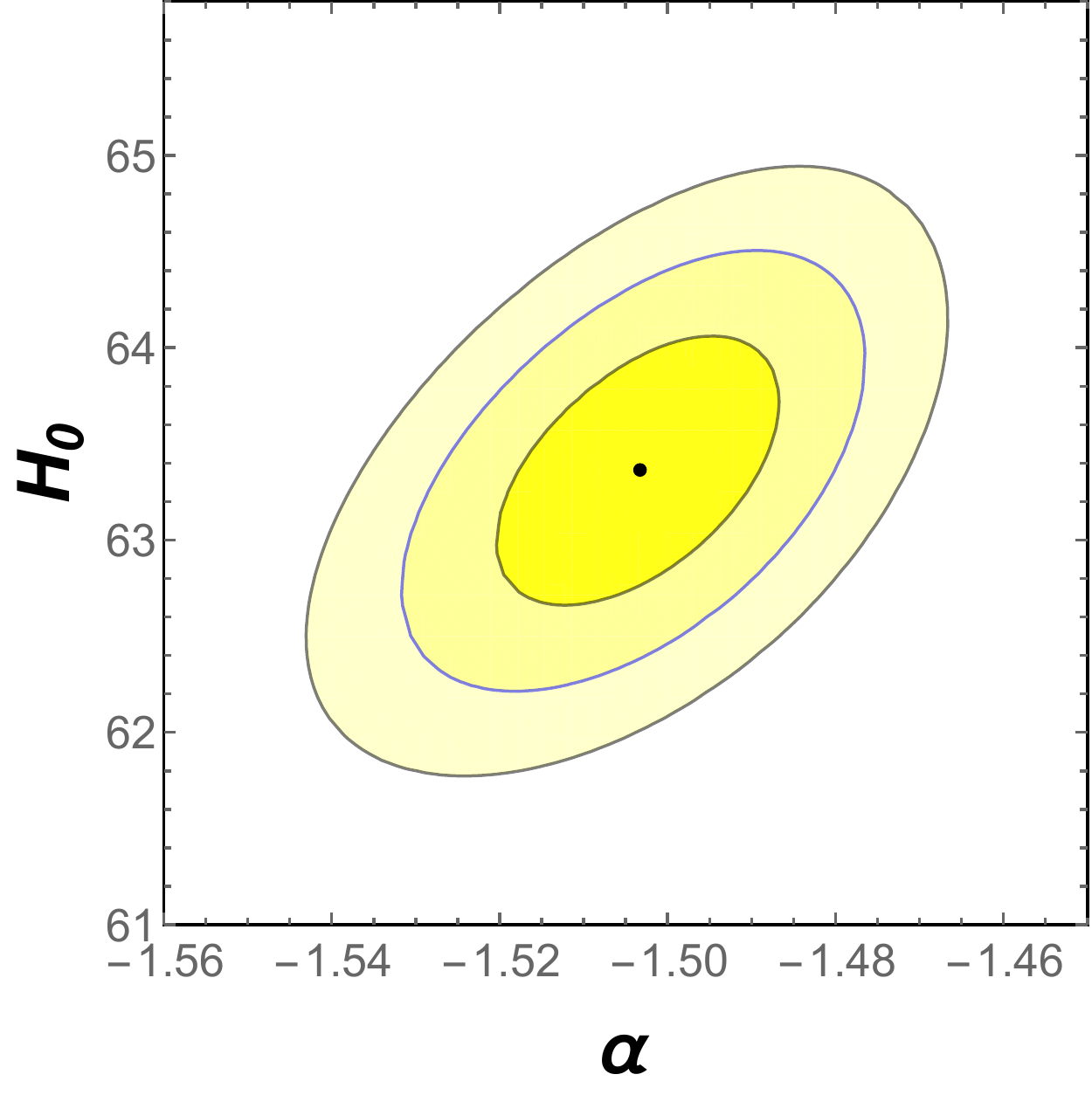} \\ 
\mbox (a) & \mbox (b)%
\end{array}%
$%
\end{center}
\caption{Figures (a) shows the maximum likelihood contours in the $\protect%
\alpha $-$H_{0}$ plane for $H(z)$ datasets only while figure (b) shows the
maximum likelihood for $H(z)$ + $SNeIa$ + $BAO$ datasets jointly. The three
contour regions shaded with dark, light shaded and ultra lightshaded in both
the plots are with errors at $1$-$\protect\sigma $, $2$-$\protect\sigma $
and $3$-$\protect\sigma $ levels. The black dots represent the best fit
values of model parameter $\protect\alpha $ and $H_{0}$ in both the plots.}
\end{figure}

\section{Results and Discussions}\label{VIII}

The manuscript communicates the phenomena of late time acceleration in the
framework of hybrid and logarithmic teleparallel gravity. To obtain the
exact solutions of the field equations, we employ a parametrization of
deceleration parameter first proposed in \cite{banerjee}. In this section,
we shall discuss the energy conditions and the cosmological viability of the
underlying teleparallel gravity models. \newline
In Section \ref{VI}, we show the temporal evolution of SEC, NEC and WEC for
both the teleparallel gravity models. Note that in order to suffice the late
time acceleration, the SEC has to violate \cite{non41to42}. This is due to
the fact that for an accelerating universe compatible with observations \cite%
{observations}, the EoS parameter $\omega \simeq -1$ \cite{planck}, and
therefore $\rho (1+3\omega )<0$ always. From Fig.\ref{f14}-\ref{f15}, one
can clearly observe that SEC violate for both the models whereas NEC and WEC
do not violate. Interestingly, SEC is also violated for curvature coupled
and minimally coupled scalar field theories \cite{non41to42}. \newline
To understand the cosmological viability of both the teleparallel gravity
models, we show in Fig. \ref{f1}, the deceleration parameter ($q$) as a
function of redshift. The plot of the deceleration parameter $q(z)$ clearly
shows that our model successfully generates late time cosmic acceleration
along with a decelerated expansion in the past for $-1>\alpha >-2$. The
deceleration parameter undergoes a signature flipping at the redshift $%
z_{tr}\simeq0.6$ for the chi-square value of $\alpha =-1.5$ which is
compatible with latest Planck measurements \cite{planck}. Our values of $%
q_{0}=-0.251355$ and $z_{tr}$ are consistent with values reported by other
authors \cite{Jaime/2019}. From Figs. \ref{f5} \& \ref{f9} the values of EoS 
$\omega $ at $z=0$ for both our models are obtained as $-0.500903$ and $%
-0.500935$ respectively. These values of $\omega $ behave in concordance
with standard cosmological model predictions (with Plank data, $\omega
_{eff}^{LCDM}\sim -0.68$ at $z=0$ as in Ref. \cite{planck}).\newline
In Fig. \ref{f2},\ref{f3},\ref{f6} and \ref{f7}, we plot the energy density
for both the models as a function of redshift. We chose the model parameters
so as to satisfy the WEC. Fulfillment of WEC ensures the cosmological
pressure has to be negative to account for negative EoS parameter and
therefore the cosmic acceleration. It is interesting to note that no known
entity has the remarkable property of negative pressure and can only be
achieved by exotic matter or by modifications to general relativity. \newline
The EoS parameter is an important cosmological parameter which has sparked
great deal of interest among cosmologists. Owing to the mysterious nature of
the cosmological entity responsible for this acceleration, various dark
energy models have been devised to suffice the observations. To investigate
the nature of the dark energy model represented by the equation \eqref{5e},
we study in Section \ref{VI}, the $\{r,s\}$ and $\{r,q\}$ plane and $Om(z)$.
We observe that the value of $\alpha $ dictates the evolution of the $r$-$s$
and $r$-$q$ trajectories. We find the model to deviate significantly from
the $\Lambda $CDM at early times. However, at late times the model is
observed to coincide with ($r=1,s=0$) and therefore consistent with $\Lambda 
$CDM cosmology. This result is further re-assured from the $r$-$q$ plane in
Fig. \ref{frq}. However, discrepancy arises from Fig. \ref{f13}, where for
none of the values of $\alpha $ we obtain a constant $Om$ which clearly does
not reflect a dark energy which is time independent. Furthermore, the nature
of dark energy represented by the equation \eqref{5e} changes from being an
Quintessence to Phantom as $\alpha $ changes from $\alpha \leq -2$ to $%
\alpha >-2$ respectively. Finally, we have discussed our obtained models in
the light of some observational datasets. The obtained model has a nice fit
to the $57$ points of Hubble datasets and the $580$ points of Union$2.1$
compilation supernovae datasets. We have used the Bayesian statistics to
find the constraints on the model parameters. The maximum likelihood
contours for the model parameter $\alpha $ and Hubble constant $H_{0}$ with
errors at $1$-$\sigma $, $2$-$\sigma $ and $3$-$\sigma $ levels in the $%
\alpha $-$H_{0}$ plane is shown separately for $H(z)$ datasets only and
joint datasets $H(z)$ + $SNeIa$ + $BAO$. The best fit constrained values of $%
\alpha $ and $H_{0}$ are found to be $\alpha =-1.497294$ \& $H_{0}=63.490604$
due for $H(z)$ datasets with $\chi _{\min }^{2}=31.333785$ and $\alpha
=-1.503260$ \& $H_{0}=63.361612$ due to $H(z)$ + $SNeIa$ + $BAO$ datasets
with $\chi _{\min }^{2}=650.312968$ respectively.

\section*{Acknowledgments}  S.M.
acknowledges Department of Science \& Technology (DST), Govt. of India, New
Delhi, for awarding Junior Research Fellowship (File No. DST/INSPIRE
Fellowship/2018/IF180676). SB thanks Biswajit Pandey for helpful discussions. PKS acknowledges CSIR, New Delhi, India for
financial support to carry out the Research project [No.03(1454)/19/EMR-II
Dt.02/08/2019]. We are very much grateful to the honorable referee
and the editor for the illuminating suggestions that have significantly
improved our work in terms of research quality and presentation.


\begin{thebibliography}{99}
\bibitem{observations} A. G. Riess et al., Astron. J. \textbf{116}, 1009
(1998); S. Perlmutter et al., Astrophys. J. \textbf{517}, 565 (1999); P.
deBernardis et al., Nature \textbf{404}, 955 (2000); S. Perlmutter et al.,
Astrophys. J. \textbf{598}, 102 (2003); M. Colless et al., Mon. Not. R.
Astron. Soc. \textbf{328}, 1039 (2001); M. Tegmark et al., Phys. Rev. D 
\textbf{69}, 103501 (2004); S. Cole et al., Mon. Not. R. Astron. Soc. 
\textbf{362}, 505 (2005); V.Springel et al., Nature (London) \textbf{440},
1137 (2006); P. A. R. Ade et al., Astron. Astrophys. \textbf{571}, A16
(2014); P. Astier et al., Astron. Astrophys. \textbf{447}, 31 (2006); A. G.
Riess et al., Astrophys. J. \textbf{659}, 98 (2007); D. N. Spergel et al.,
Astrophys. J. Suppl. Ser. \textbf{148}, 175 (2003); H. V. Peiris et al.,
Astrophys. J. Suppl. Ser. \textbf{148}, 213 (2003); D. N. Spergel et al.,
Astrophys. J. Suppl. Ser. \textbf{170}, 377 (2007); E. Komatsu et al.,
arXiv:0803.0547.

\bibitem{alternate} B. Ratra, P. J. E Peebles, Phys. Rev. D \textbf{37},
3406 (1988); R. R. Caldwell et al., Phys. Rev. Lett. \textbf{80}, 1582 1988;
C. Armendariz-Picon et al., Phys. Rev. D \textbf{63}, 103510 (2001); T.
Buchert, Gen. Relativ. Gravit. \textbf{32}, 105 (2000); P. Hunt, S. Sarkar,
Mon. Not. R. Astron. Soc. \textbf{401}, 547 (2010); K. Tomita, Mon. Not. R.
Astron. Soc. \textbf{326}, 287 (2001); B. Pandey, Mon. Not. R. Astron. Soc. 
\textbf{485}, L73 (2019); B. Pandey, Mon. Not. R. Astron. Soc. \textbf{471},
L77 (2017); K. A. Milton, Gravit. Cosmol.\textbf{\ 9}, 66 (2003); D. Easson
et al., Phys. Lett. B \textbf{696}, 273 (2011); D. Pav\'on , N. Radicella,
Gen. Relativ. Gravit. \textbf{45}, 63 (2013); N. Radicella, D. Pav\'on, Gen.
Relativ. Gravit. \textbf{44}, 685 (2012).

\bibitem{cosmoft} A. Paliathanasis et al., Phys. Rev. D \textbf{94}, 023525
(2016).

\bibitem{review} S. Nojiri et al., Phys. Rept. \textbf{692}, 1 (2017).
\bibitem{snehasish}  P.K. Sahoo and S. Bhattacharjee, New Astronomy, \textbf{77}, 101351 (2020); R. Zaregonbadi, et al., Phys. Rev. D \textbf{94}, 084052 (2016); G. Sun and Y.-C. Huang, Int. J. Mod. Phys. D, \textbf{25}, 1650038 (2016); F. Rocha et al. arXiv:1911.08894 (2019); S.I. dos Santos, G.A. Carvalho, P.H.R.S. Moraes, C.H. Lenzi and M. Malheiro, Eur. Phys. J. Plus, \textbf{134}, 398 (2019); P.H.R.S. Moraes, J.D.V. Arban˜il and M. Malheiro, J. Cosm. Astrop. Phys. \textbf{06}, 005 (2016); P.H.R.S. Moraes and P.K. Sahoo, Eur. Phys. J. C \textbf{79}, 677 (2019); E. Elizalde and M. Khurshudyan, Phys. Rev. D, \textbf{99}, 024051 (2019); P.H.R.S. Moraes, W. de Paula and R.A.C. Correa, Int. J. Mod. Phys. D, \textbf{28}, 1950098 (2019); E. Elizalde and M. Khurshudyan, Phys. Rev. D, \textbf{98}, 123525 (2018); P.H.R.S. Moraes and P.K. Sahoo, Phys. Rev. D, \textbf{97}, 024007 (2018); P.K. Sahoo, P.H.R.S. Moraes and P. Sahoo, Eur. Phys. J. C, \textbf{78}, 46 (2018); P.K. Sahoo, P.H.R.S. Moraes, P. Sahoo and G. Ribeiro, Int. J. Mod. Phys. D, \textbf{27}, 1950004 (2018); P.H.R.S. Moraes and P.K. Sahoo, Phys. Rev. D, \textbf{96}, 044038 (2017); P.H.R.S. Moraes, R.A.C. Correa and R.V. Lobato, J. Cosm. Astrop. Phys., \textbf{07}, 029 (2017); T. Azizi, Int. J. Theor. Phys. \textbf{52}, 3486 (2013); M. Sharif and A. Siddiqa, Gen. Rel. Grav., \textbf{51}, 74 (2019);  M.E.S. Alves, P.H.R.S. Moraes, J.C.N. de Araujo and M. Malheiro, Phys. Rev. D, \textbf{94}, 024032 (2016); P.K. Sahoo and S. Bhattacharjee, Int. J. Theor. Phys, DOI: 10.1007/s10773-020-04414-3 [arXiv: 1907.13460]; S. Bhattacharjee and P. K. Sahoo, Eur. Phys. J. C. \textbf{80}, 289 (2020) [arXiv: 2002.11483]; 
 P. Sahoo et al, Mod. Phys. Lett. A, DOI: 10.1142/S0217732320500959 [arXiv: 1907.08682] ;S. Bhattacharjee and P. K. Sahoo, Phys. Dark. universe. \textbf{28}, 100537 (2020) [arXiv: 2003.14211]; S. Bhattacharjee and P. K. Sahoo, Eur. Phys. J. Plus, \textbf{135}, 86 (2020) [arXiv:2001.06569]; S. Bhattacharjee and P. K. Sahoo, Eur. Phys. J. Plus, DOI: 10.1140/epjp/s13360-020-00361-4.
\bibitem{cosmo7to9} G. R. Bengochea, R. Ferraro, Phys. Rev. D \textbf{79},
124019 (2009); R. Ferraro, F. Fiorini, Phys. Rev. D \textbf{75}, 084031
(2007); E. V. Linder, Phys. Rev. D \textbf{81}, 127301 (2010).

\bibitem{reviewft} Y. F. Cai et al., Rept. Prog. Phys. \textbf{79}, no.4,
106901 (2016).

\bibitem{cosmo10to12} K. Hayashi, T. Shirafuji, Phys. Rev. D \textbf{19},
3524 (1979); M. Tsamparlis, Phys. Lett. A \textbf{75}, 27 (1979); H. I.
Arcos, J.G. Pereira, Int. J. Mod. Phys. D \textbf{13}, (2004) 2193

\bibitem{cosmo14} A. Einstein 1928, Sitz. Preuss. Akad. Wiss. p. 217; ibid
p. 224 [Translated by A. Unzicker and T. Case, (preprint: arXiv:
physics/0503046)]

\bibitem{cosmo21} M. E. Rodrigues et al., Astroph. Space Sci. \textbf{357},
129 (2015)

\bibitem{cosmo22to23} A. Paliathanasis et al., Phys. Rev. D \textbf{89},
104042 (2014); S. Capozziello et al., JHEP \textbf{89}, 02 039 (2013).

\bibitem{cosmo19to20} K. Atazadeh, F. Darabi, Eur. Phys. J. C \textbf{72},
2016 (2012); S. Basilakos et al., Phys. Rev. D \textbf{88}, 103526 (2013).

\bibitem{inflation} K. Bamba et al., Phys. Rev. D \textbf{94}, 083513 (2016).

\bibitem{late} G. R. Bengochea, R. Ferraro, Phys. Rev. D \textbf{79}, 124019
(2009); K. Bamba et al., JCAP \textbf{01}, 021 (2011); R. Myrzakulov, Eur.
Phys. J. C \textbf{71}, 1752 (2011).

\bibitem{bounce} Y.F. Cai et al., Class. Quant. Grav. \textbf{28}, 215011
(2011); J. de Haro, J. Amor\'os, J. Phys. Conf. Ser. \textbf{600}, 012024
(2015); J. de Haro , J. Amor\'os , PoS FFP14, 163 (2016); W. El Hanafy, G.
G. L. Nashed, Int. J. Mod. Phys. D \textbf{26}, 1750154 (2017).

\bibitem{bhatti/2017} M. Zaeem-ul-HaqBhatti, Z.Yousaf, Sonia Hanif, Phys.
Dark Univ.,\textbf{16}, 34 (2017).

\bibitem{bhatti/2017a} M. Zaeem-ul-Haq Bhatti, Z.Yousaf, Sonia Hanif, Mod.
Phys. Lett. A, \textbf{32}, 1750042 (2017).

\bibitem{pacif} S. K. J. Pacif et al., Int. J. Geom. Meth. Mod. Phys. 
\textbf{14}, 1750111 (2017).

\bibitem{para} M. Chevallier, D. Polarski, Int. J. Mod. Phys. D \textbf{10},
213 (2001); Mod. Phys. Lett. A. \textbf{35}, 2050011 (2020); S. del Campo et
al., Phys. Rev. D \textbf{86}, 083509 (2012); A. A. Mamon, K.Bamba, Eur.
Phys. J. C \textbf{78}, 862 (2018).

\bibitem{banerjee} N. Banerjee, S. Das, Gen. Relativ. Gravit. \textbf{37},
1695 (2005).

\bibitem{olive21} S. Pan, A.Mukherjee, N.Banerjee, Mon. Not. R. Astron. Soc. 
\textbf{477}, 1, 1189 (2018).

\bibitem{sahni} V. Sahni, T. D. Saini, A. A. Starobinsky, U. Alam: JETP
Lett. \textbf{77}, 201 (2003); U. Alam et al., Mon. Not. R. Astron. Soc. 
\textbf{344}, 1057 (2003).

\bibitem{sahni17to18} J. Albert et al. [SNAP Collaboration]: arXiv:0507458;
J. Albert et al. [SNAP Collaboration]: arXiv:0507459.

\bibitem{Omsahni} V. Sahni et al., Phys. Rev. D. \textbf{78}, 103502 (2008);
M. Shahalam1, Sasha Sami, Abhineet Agarwal, Mon. Not. Roy. Astron. Soc. 
\textbf{448} 2948 (2015); Abhineet Agarwal et al., Int. J. Mod. Phys. D%
\textbf{28} 1950083 (2019).

\bibitem{sahoo} P. H. R. S. Moraes, P. K. Sahoo, Eur. Phys. J. C. \textbf{77}%
, 480 (2017).

\bibitem{non39} R. M. Wald, General relativity (University of Chicago Press,
Chicago, 1984)

\bibitem{energy} D. Liu, M. J. Reboucas, Phys.Rev. D. \textbf{86}, 083515
(2012); T. Azizi, M. Gorjizadeh, EPL \textbf{117}, 6 (2017);M. Zubair, Saira
Waheed, Astrophys Space Sci. \textbf{355}, 361 (2015).

\bibitem{sharov} G. S. Sharov, V. O. Vasiliev, Mathematical Modelling and
Geometry, Vol. 6, No 1, 1 (2018).

\bibitem{Hz1} C. H. Chuang, Y. Wang, Mon. Not. Roy. Astron. Soc. \textbf{435}%
, 255 (2013); C-H Chuang C-H. et al., Mon. Not. Roy. Astron. Soc. \textbf{433%
}, 3559 (2013); A. Font-Ribera et al., J. Cosmol. Astropart. Phys. \textbf{05%
}, 027 (2014); T. Delubac et al., Astron. Astrophys. \textbf{574}, id. A59,
17 (2015); L. Anderson et al., Mon. Not. Roy. Astron. Soc. \textbf{441}, 24
(2014); Y. Wang et al., Mon. Not. Roy. Astron. Soc. \textbf{469}, 3762
(2017); E. Gazta naga et al. Mon. Not. Roy. Astron. Soc. \textbf{399}, 1663
(2009); C. Blake et al., Mon. Not. Roy. Astron. Soc. \textbf{425}, 405
(2012); N. G. Busca et al., Astro. Astrophys. \textbf{552}, A96 (2013); A.
Oka et al., Mon. Not. Roy. Astron. Soc. \textbf{439}, 2515 (2014); S. Alam
et al., Mon. Not. Roy. Astron. Soc. \textbf{470}, 2617 (2017); J. E.
Bautista et al., Astron. Astrophys. \textbf{603}, id. A12, 23 (2017).

\bibitem{Hz2} J. Simon, L. Verde, R. Jimenez, Phys. Rev. D \textbf{71},
123001 (2005); D. Stern et al., J. Cosmol. Astropart. Phys. \textbf{02}, 008
(2010); C. Zhang et al., Research in Astron. and Astrop. \textbf{14}, 1221
(2014); M. Moresco et al., J. Cosmol. Astropart. Phys. \textbf{8}, 006
(2012); M. Moresco, Mon. Not. Roy. Astron. Soc.: Letters., \textbf{450}, L16
(2015); M. Moresco M. et al., J. Cosmol. Astropart. Phys. \textbf{05}, 014
(2016); A. L. Ratsimbazafy et al., Mon. Not. Roy. Astron. Soc. \textbf{467},
3239 (2017)

\bibitem{SNeIa} N. Suzuki \textit{et al.}, Astrophys. J., \textbf{746}
(2012) 85

\bibitem{padn} N. Padmanabhan, X. Xu, D. J. Eisenstein, R. Scalzo, A. J.
Cuesta, K. T. Mehta et al., Mon. Not. Roy. Astron. Soc. \textbf{427} (2012)
2132

\bibitem{6df} F. Beutler, C. Blake, M. Colless, D. H. Jones, L.
Staveley-Smith, L. Campbell et al., Mon. Not. Roy. Astron. Soc. \textbf{416}
(2011) 3017

\bibitem{boss} BOSS collaboration, L. Anderson et al., Mon. Not. Roy.
Astron. Soc. \textbf{441} (2014) 24

\bibitem{wig} C. Blake et al., Mon. Not. Roy. Astron. Soc. \textbf{425}
(2012) 405

\bibitem{adep} Ade P A R et al., Planck 2015 results. XIII. Cosmological
parameters, Preprint arXiv:1502.01589 (2015)

\bibitem{waga} M. Vargas dos Santos, Ribamar R. R. Reis, J. Cosm. Astropart.
Phys., \textbf{1602} (2016) 066

\bibitem{gio} R. Giostri, M. V. d. Santos, I. Waga, R. R. R. Reis, M. O.
Calvao and B. L. Lago, JCAP \textbf{1203} (2012) 027

\bibitem{hing} G. Hinshaw et al., Astrophys. J. Suppl., \textbf{208} (2013)
19

\bibitem{non41to42} M. Visser, C. Barcelo. arXiv:gr-qc/0001099; C. Barcelo,
M. Visser, Int. J. Mod. Phys. D \textbf{11}, 1553 (2002). arXiv:gr-qc/0205066

\bibitem{planck} Planck Collaboration: arXiv:1807.06209.

\bibitem{Jaime/2019} B. S. Haridasu, V. V. Lukovic, M. Moresco, N. Vittorio,
J. Cosmol. Astropart. Phys. \textbf{10}, 015 (2018); J. R. Garza et al.,
Eur. Phys. J. C \textbf{79}, 890 (2019); Hai-Nan Lin, Xin Li, Li Tang, Chin.
Phys. C \textbf{43}, 075101 (2019); J. F. Jesus, R. Valentim, A. A. Escobal,
S. H. Pereira, arXiv: 1909.00090. 
%\bibitem{Perlmutter/1999} S. Perlmutter et al., Astrophys. J. \textbf{517}, 565 (1999).
%\bibitem{Riess/1998} A. G. Riess et al., Astron. J. \textbf{116}, 1009 (1998).

%\bibitem{stern/2010} D. Stern et al., J. Cosm. Astropart. Phys. \textbf{1002}, 008 (2008).
%\bibitem{Santos/2006}M. V. dos Santos, R. R. R. Reis, I. Waga, J. Cosmol. Astropart. Phys. \textbf{2016}, 066 (2016).
%\bibitem{Gong/2006} Y. Gong, A. Wang, Phys. Rev. D \textbf{73}, 083506 (2006).
%\bibitem{Sahni/2003}V. Sahni, T.D. Saini, A. A. Starobinsky, U. Alam, JETP Lett. \textbf{77},
%201 (2003). 
%\bibitem{Alam/2003} U. Alam, V. Sahni, T.D. Saini, A. A. Starobinsky, MNRAS \textbf{344},
%1057 (2003). 
%\bibitem{Zhai/2013} Z.-X. Zhai, et al., Phys. Lett. B \textbf{727}, 8 (2013).
%\bibitem{Mukherjee/2016} M. Mukherjee, N. Banarjee, Phys. Pev. D \textbf{93}, 043002 (2016).
%\bibitem{Mamon/2018} A.A. Mamon, K. Bamba, Eur. Phys. J. C \textbf{78}, 862 (2018).
%\bibitem{Capozziello/2020} S. Capozziello, R. D'Agostino, O. Luongo, arXiv:2003.09341 [astro-ph.CO]
%\bibitem{farooq/2013} O. Farooq, B. Ratra, Astrophys. J. \textbf{766}, L7 (2013).
%\bibitem{bamba/2012} K. Bamba, R. Myrzakulov, S. Nojiri, S. D. Odintsov, Phys. Rev. D \textbf{85}, 104036 (2012).
%\bibitem{Moraes/2019} P. H. R. S. Moraes, P. K. Sahoo, S. K. J. Pacif, 	arXiv:1905.00417 [gr-qc].
%\bibitem{Sahni/2008} V. Sahni et al., Phys. Rev. D \textbf{78}, 103502 (2008).
%\bibitem{Zunckel/2008} C. Zunckel, C. Clarkson, Phys. Rev. Lett. \textbf{101}, 181301 (2008).
%\bibitem{Shahalam/2015} M. Shahalam, S. Sami, A. Agarwal, Mon. Not. R. Astron. Soc.
%\textbf{448}, 2948 (2015).
%\bibitem{Carroll/2004}S. Carroll, \textit{Spacetime and Geometry: An Introduction to General
%Relativity} (Addison Wesley, Boston, 2004).
%\bibitem{Santos/2005} J. Santos, J. S. Alcaniz, Phys. Lett. B \textbf{619}, 11 (2005).
%\bibitem{Santos/2007} J. Santos et al., Phys. Rev. D \textbf{76}, 043519 (2007).
%\bibitem{Sen/2008} A. A. Sen, R. J. Scherrer, Phys. Lett. B \textbf{659}, 457 (2008).
%\bibitem{Santos/2010} J. Santos et al., Int. J. Mod. Phys. D \textbf{19}, 1315 (2010).
%\bibitem{Bertolami/2009} O. Bertolami, M. C. Sequeira, Phys. Rev. D \textbf{79}, 104010 (2009).
%\bibitem{Nojiri/2008} S. Nojiri et al., Prog. Theor. Phys. Suppl. \textbf{172}, 81 (2008).
%\bibitem{Garcia/2011} N. M. Garcia et al., Phys. Rev. D \textbf{83}, 104032 (2011).
%\bibitem{Banijamali/2012} A. Banijamali et al., Astrophys. Space Sci. \textbf{338}, 327 (2012).
\end{thebibliography}
\end{document}